\documentclass[fleqn,10pt]{wlscirep}
\usepackage{siunitx}
\usepackage{bm}
\usepackage{pgfplots}
\pgfplotsset{compat=newest}
\usepackage{tikz}
\usetikzlibrary{positioning,pgfplots.groupplots,matrix}
\usetikzlibrary{fit}
\usepackage[space]{grffile}
\usepackage{tabularx}
\usepackage{setspace}\usepackage{threeparttable}

\newcommand{\zb}[1]{\boldsymbol{#1}}

\definecolor{c1}{RGB}{166,206,227}
\definecolor{c2}{RGB}{31,120,180}
\definecolor{c3}{RGB}{178,223,138}
\definecolor{c4}{RGB}{51,160,44}
\definecolor{c5}{RGB}{251,154,153}
\definecolor{c6}{RGB}{227,26,28}

\title{Simultaneous Magnetic Particle Imaging and Navigation of large superparamagnetic nanoparticles in bifurcation flow experiments}
\author[1,2,*]{Florian Griese}
\author[1,2]{Tobias Knopp}
\author[4]{Cordula Gruettner}
\author[1,2]{Florian Thieben}
\author[4]{Knut Müller}
\author[5]{Sonja Loges}
\author[3,$\ddagger$]{Peter Ludewig}
\author[1,2,$\ddagger$]{Nadine Gdaniec}
\affil[1]{University Medical Center Hamburg-Eppendorf, Section for Biomedical Imaging, Hamburg,  Germany}
\affil[2]{Hamburg University of Technology, Institute for Biomedical Imaging, Hamburg, Germany}
\affil[3]{University Medical Center Hamburg-Eppendorf, Department of Neurology, Hamburg, Germany}
\affil[4]{micromod Partikeltechnologie GmbH, Rostock, Germany}
\affil[5]{University Medical Center Hamburg-Eppendorf, Department of Oncology, Hematology and Bone Marrow Transplantation with Section Pneumology, Hubertus Wald Tumorzentrum, Hubertus Wald Comprehensive Cancer Center Hamburg, Hamburg, Germany}
\affil[6]{University Medical Center Hamburg-Eppendorf, Department of Tumor Biology, Center of Experimental Medicine, Hamburg, Germany}

\affil[*]{f.griese@uke.de}
\affil[$\ddagger$]{Contributed equally}


\begin{abstract}
Magnetic Particle Imaging (MPI) has been successfully used to visualize the distribution of superparamagnetic nanoparticles within 3D volumes with high sensitivity in real time. Since the magnetic field topology of MPI scanners is well suited for applying magnetic forces on particles and micron-sized ferromagnetic devices, MPI has been recently used to navigate micron-sized particles and micron-sized swimmers. In this work, we analyze the magnetophoretic mobility and the imaging performance of two different particle types for Magnetic Particle Imaging/Navigation (MPIN). MPIN constantly switches between imaging and magnetic modes, enabling quasi-simultaneous navigation and imaging of particles. We determine the limiting flow velocity to be \SI{8.18}{\milli\liter\per\second} using a flow bifurcation experiment, that allows all particles to flow only through one branch of the bifurcation. Furthermore, we have succeeded in navigating the particles through the branch of a bifurcation phantom narrowed by either 60\% or 100\% stenosis, while imaging their accumulation on the stenosis. The particles in combination with therapeutic substances have a high potential for targeted drug delivery and could help to reduce the dose and improve the efficacy of the drug, e.g. for specific tumor therapy and ischemic stroke therapy.
\end{abstract}
\begin{document}

\flushbottom
\maketitle

\thispagestyle{empty}


\section{Introduction}
Magnetic Particle Imaging (MPI) uses non-linear magnetization characteristics to spatially resolve the distribution of superparamagnetic nanoparticles in 3D at high temporal resolution.\cite{gleich_tomographic_2005,knopp_online_2016} The highest sensitivity is achieved with particles in the magnetic core size range of \SI{15}{\nano\meter} to \SI{30}{\nano\meter}.\cite{ferguson_magnetic_2015,ferguson_optimization_2009} Due to different signal characteristics for different particle types, it is even possible to distinguish between different particle types with the multi-contrast approach introduced by Rahmer et al.\cite{Rahmer2015}
Lately, MPI has been utilized to track balloon catheters for stenosis clearing\cite{Salamon2016a,Herz2018} and to navigate magnetically coated catheters through difficult bifurcations\cite{Rahmer2017}. To this end, a soft-magnetic sphere has been attached to the catheter. With both possibilities it is feasible to simultaneously image the catheter's position with the coated magnetic particles and steer the catheter in any direction by applying a magnetic force on the sphere.\cite{Rahmer2017} It has already been shown that micron-sized devices with soft-magnetic spheres attached can be tracked by the imaging capabilities, and moved to target areas by the force capabilities, of MPI.\cite{Nothnagel2016} A similar principle is applied with rotational magnetic fields to selectively control helical micro-devices. In terms of application, these screw-like devices might be used to unscrew a radioactive source out of its shielding close to a target region e.g.~for local cancer treatment.\cite{Rahmereaal2845} The same principle has been demonstrated on a human scale for multiple screws by Rahmer et al.\cite{rahmer_remote_2018} Bakenecker et al.\cite{BAKENECKER2018} have also utilized rotational focus fields to control the actuation of magnetically coated swimmers and have moved them through bifurcations while simultaneously imaging and magnetically navigating them. The ability to move particles with MPI for targeted drug delivery has also been presented. \cite{le_real-time_2017,mahmood_novel_2015,zhang_development_2017,kuboyabu_usefulness_2016,bente_biohybrid_2018}
Griese et al.\cite{griese_imaging_2018} have demonstrated how to simultaneously move and image micron-sized magnetic particles with the Magnetic Particle Imaging/Navigation (MPIN) method with a temporal resolution of \SI{2.9}{\hertz}. Micron-sized particles have been used to target endothelial cells of the central nervous system to identify over-expressed surface proteins such as ICAM-1. The over-expression caused by neurological diseases can be imaged by Molecular Magnetic Resonance Imaging.\cite{gauberti_molecular_2018} The micron-sized particles do not cross the blood brain barrier and the enhanced permeability retention effect occurring with smaller particles is insignificant. 

\begin{figure}[hbt]
  \centering
    \includegraphics[width=10.0cm]{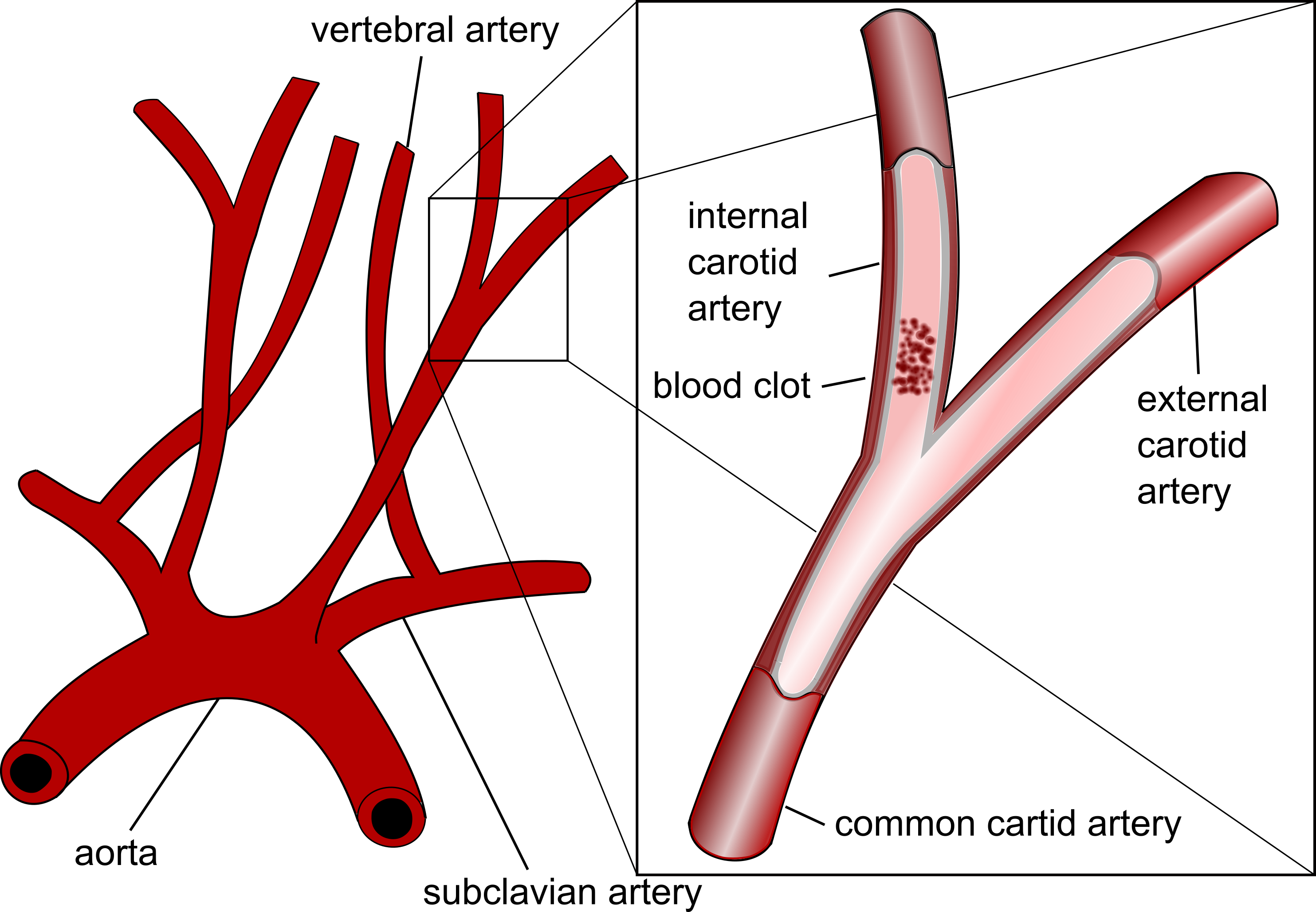}
    \caption{A blood clot causing a stenosis in internal carotid artery. The blood flow at the bifurcation of the internal and external carotid artery goes mainly through the unblocked external artery making it hard for therapeutic substances to take effect.}
    \label{fig:CarotisStenosis}
\end{figure}

The ability to navigate and image magnetic particles with MPIN qualifies for the following potential application scenarios. 
In the case of targeted drug delivery, magnetic particles can be attached to therapeutic substances (e.g. fibrinolytic medications) and be injected into an organism. In the case of acute stroke, the MPIN method could navigate and concentrate the medicated particles locally in the vascular territory of the vessel occlusion (see Fig.~\ref{fig:CarotisStenosis}), thus enhancing the disintegration of the blood clot. A normal injection with thrombolytic medications without magnetic particles can result in ineffective dissolving of the blood clot because the blood flows mainly through the unblocked branch of the bifurcation between external carotid artery and internal carotid artery. Normally, the blood clot is then cleared using a catheter in an invasive procedure.
Therefore, the usage of therapeutic functionalized magnetic particles could help reduce the dose of the medication or even make an invasive intervention superfluous. 

In this work, we investigate two different types of particles with different magnetic core diameters and determine which type satisfies the best compromise in terms of magnetic manipulability and best imaging performance for MPI. To quantify the navigation characteristics of the particles outside the MPI system, we analyze the magnetophoretic mobility. Furthermore, we perform measurements with a Magnetic Particle Spectrometer (MPS) to identify the particles' imaging characteristics. Finally, we use inflow bifurcation experiments to investigate the flow velocity, up to which it is possible to navigate the particles to one side of the bifurcation, even with a stenosis in one branch of the bifurcation phantom. 

\section{Theory}
\subsection{Magnetic Navigation of Particles}
The magnetic force on an isotropic suspension of magnetic particles in a magnetic gradient field can be determined to be
\begin{eqnarray}\label{eq:ForceZ1}
\zb F_{m}=(\zb m\cdot \nabla)\zb B=\frac{V_m \Delta \chi}{\mu_0}(\zb B \cdot \nabla)\zb B=V_m \Delta \chi\nabla(\frac{\lVert{\zb B}\rVert^2}{2\mu_0})=\frac{4}{3}\pi r_m^3 \Delta \chi\nabla(\frac{\lVert {\zb  B}\rVert^2}{2\mu_0}) 
\end{eqnarray}
with the dipole moment $\zb m$, the magnetic flux density $\zb B$, the vacuum permeability $\mu_0$, the difference in magnetic susceptibility $\Delta \chi$, the volume $V_m$, and $r_m$ the magnetic core radius of the particles.\cite{krishnan_fundamentals_2016}[p.592]

The magnitude of the magnetic selection field of an MPI field-free point scanner is linear, increasing with distance to the field-free point (FFP). In one direction, e.g. the $y$-direction, this results in 
\begin{equation}
    B_y=\mu_0  H_y = \mu_0 G_y y
\end{equation}
with the gradient strength $G_y = \frac{\partial H_y}{\partial y}$, the magnetic field $\zb H$, and assuming the origin of coordinates to be the field-free point.
Thus, the magnetic force induced by the gradient field of an MPI scanner on a magnetic particle\cite{Nothnagel2016} in the $y$-direction can be expressed as
\begin{eqnarray}\label{eq:ForceZ2}
\zb F_{m}=\frac{4}{3}\pi r_m^3 \Delta \chi\nabla(\frac{\lVert{\zb B}\rVert^2}{2\mu_0})=\frac{2\mu_0}{3}\pi r_m^3 \Delta \chi\nabla(G_y^2y^2)=\frac{4\mu_0}{3}\pi r_m^3 \Delta \chi G_y^2y,
\end{eqnarray}
with $y$ being the distance to the FFP. Opposed to the magnetic force, a drag force acts on a magnetic particle if the particle is moving within a liquid such as water or blood.
The drag force, for simplicity given here in the $y$-direction, is defined by
\begin{eqnarray}\label{eq:dragforce}
F_{d,y}(v)=6\pi\eta_{water} r_{h} \Delta v_{m,y}
\end{eqnarray}
with $\eta_{water}$ being the fluid viscosity, $r_{h}$ being the hydrodynamic radius of the particles, and $\Delta v_{m,y}$ being the velocity difference of the particles to the flowing liquid.

Taking a look at the motion of particles in the horizontal plane in the case of balancing drag force and magnetic force $F_{m,y}=F_{d,y}$, one can obtain the equilibrium difference velocity of the particles and the fluid using
\begin{eqnarray}
\Delta v=\frac{\Delta \chi r_{m}^{3}}{9 \mu_0 \eta r_{h}}\nabla B_{0}^2=\xi\nabla B_{0}^2=\xi\nabla (\mu_{0}H(y))^2=\xi\frac{\partial}{\partial y} (\mu_{0}^2(G_{y} y)^2)= \xi\mu_{0}^2G_{y}^2 y
\label{eq:velocityDifference}
\end{eqnarray}
with the magnetophoretic mobility\cite{krishnan_fundamentals_2016}[p.593],\cite{kara_e._mccloskey_magnetic_2003},\cite{zhou_magnetic_2016} defined as
\begin{eqnarray}
\xi=\frac{\Delta v}{\nabla B_0^2}.
\end{eqnarray}
For the described magnetic carrier, according to equation~\eqref{eq:velocityDifference}, this results in 
\begin{eqnarray}
\xi=\frac{\Delta \chi r_{m}^{3}}{9 \mu_{0} \eta r_{h}}
\end{eqnarray}
with unit [$\frac{m^2 s^3 A^2 }{kg^2}$]. The magnetophoretic mobility can be calculated based on explicit knowledge of characteristic parameters of the particles or based on their velocity in a magnetic gradient field inside a fluid. Following the paths of individual cells in a clearly defined magnetic gradient field is challenging but it is possible to determine indirect measures of the magnetophoretic mobility. For our purpose, a relative measure for the mobility is sufficient. For a given set of particles, the particles should be ordered according to their navigation capabilities. 

\subsubsection{Navigation characterization using separation apparatus}
In magnetic cell separation procedures an apparatus is used to separate cells from a surrounding medium with the help of magnetic fields. The duration required for the separation in these procedures is inversely proportional to the magnetophoretic mobility. Thus, by comparing these separation times for different particles, a classification regarding their suitability for navigation is possible. The procedure is described in the following.

The work of Andreu et~al.\cite{andreu_simple_2011} describes an apparatus for magnetic separation in detail and derives an analytical model for the separation procedure and the required separation times. The separation apparatus is composed of a radial magnetic gradient as used by De Las Cuevas et al. and Benelmekki et al. \cite{cuevas,benelmekki} A homogeneous dispersion of magnetic particles is placed inside a cylindrical cavity of radius $L$. A magnetic gradient field is imposed on the cavity with a uniform magnetic gradient pointing towards the walls of the vessel. The magnetic gradient enforces the particles to move radially towards the vessel wall, which is the final stage of the process, resulting in an inhomogeneous dispersion of particles. The magnetic particles are accumulated at the vessel wall while the center of the cavity is filled with the remaining liquid. The separation process is completed by removing the liquid from the center of the cavity.
A light source and a detector are additionally used to determine the turbidity of the solution over time. The turbidity $T$ at time $t$ can be described by
\begin{equation}
        T = T_\infty+\frac{T_0-T_{\infty}}{1+(\frac{t}{t_{50}})^2}
\end{equation}
for mono-sized Langevin particles with the initial turbidity $T_0$, the final turbidity $T_\infty$, and the half separation time $t_{50}$.\cite{witte_particle_2017} According to Andreu et al.\cite{andreu_simple_2011} the half separation time for mono-sized Langevin particles is given by
\begin{equation}\label{eq:HalfTime}
    t_{50}=\left(1-\frac{1}{\sqrt{2}}\right)\frac{L}{v_s},
\end{equation}
with the saturation velocity $v_s$ of the particles at magnetic saturation.
In equation~\eqref{eq:HalfTime} it is assumed that the field strength at the boundary is much higher than the required magnetic field to drive particles into magnetic saturation with saturation magnetization $M_s$. By replacing one $B$ in equation~\eqref{eq:velocityDifference} with $B=\frac{\mu_0 M}{\Delta \chi}$ and $M=M_s$, the velocity at saturation $v_s$ is described by 
\begin{equation}\label{eq:saturationVelocity}
    v_s= \xi \frac{M_s \mu_0}{\Delta \chi}\nabla B_0= \xi \frac{M_s \mu_0^2}{\Delta \chi} G_y 
\end{equation}
containing the magnetophoretic mobility $\xi$.
From equation~\eqref{eq:HalfTime} and equation~\eqref{eq:saturationVelocity} it can be seen that the half separation time $t_{50}$ and the magnetophoretic mobility are inversely proportional to each other. Thus, comparing particles based on the half-separation time gives a hint of the magnetic navigation capabilities relative to each other.  
\subsection{Simultaneous Magnetic Particle Imaging/Navigation Principle}
\begin{figure}[hbt]
  \centering
    \includegraphics[width=16.0cm]{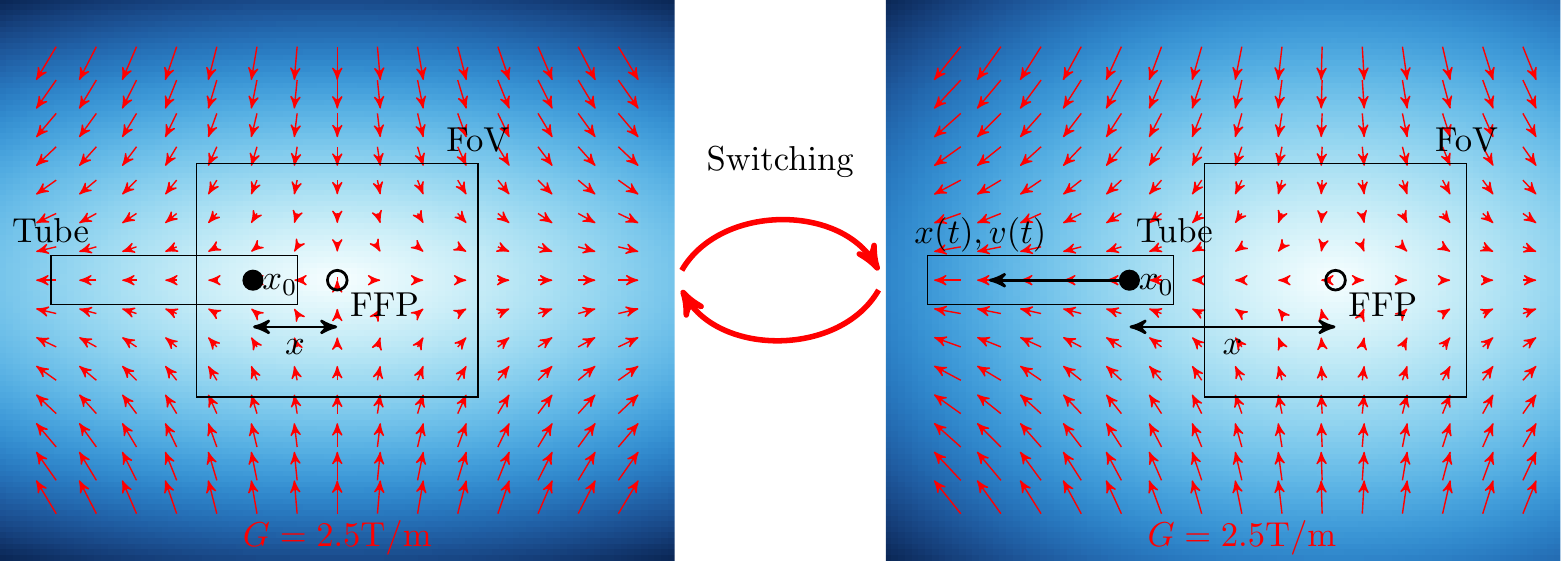}
    \caption{(Left) Illustration of the FFP position during imaging mode where the magnetic particles (indicated by a black dot) are within the Field of View (FoV). (Right) Visualization of the FFP position during the magnetic force mode where the magnetic particles are not within the FoV but, due to the large distance from the FFP, the particles are moved by the magnetic force. Imaging mode and magnetic force mode cannot be performed in parallel and have to be executed in an alternating manner. }
    \label{fig:FPPImagingForce}
\end{figure}

The preclinical MPI scanner considered in this work (Bruker Biospin MRI GmbH, Ettlingen, Germany) has one pair of focus field  coils for each direction $x$, $y$, and $z$.\cite{knopp_joint_2015} These focus fields are originally designed to perform multi-patch MPI to enlarge the field of view.\cite{gleich_fast_2010,rahmer_results_2011} Currently, the MPI scanner used in this work is capable of switching between two focus field positions with a frequency of \SI{2.9}{\hertz}. The distance from the FFP determines the magnetic force as described by equation~\eqref{eq:ForceZ2}. Thus, the focus fields can be used in the MPIN method to induce a magnetic force on the particles by appropriately choosing the position of the FFP. The principle of switching between imaging mode and navigation mode can be seen in Fig.~\ref{fig:FPPImagingForce}. 
Since the magnetic force acts only during the navigation time span, equation~\eqref{eq:ForceZ2} has to be extended with a switching function $s(t)$ to the following equation
\begin{eqnarray}\label{eq:magneticForceSwitching}
F_{m,y}(y,t)=\frac{4}{3}\pi\mu_0 r_m^3 \Delta \chi G_y^2y s(t) \quad \text{with} ~~ s(t)=\begin{cases}
1 & \text{if} \quad t  \in [T(i-1)\varphi+Ti\zeta,T i(\varphi+\zeta)) \\
0 & \text{else}
\end{cases} ~ i=1,\dots,n ~~ \text{and} ~  T=\SI{0.021}{\second}
\end{eqnarray}
with the number of imaging drive field cycles (DF-cycles) $\zeta$ and navigation DF-cycles $\varphi$. $\mu_0$ is the vacuum permeability, $G_y$ is the gradient in the $y$-direction, $r_m$ is the magnetic core radius of the particles, $T$ is the time span of one acquisition cycle and $y$ is the distance to the FFP. Equation~\eqref{eq:magneticForceSwitching} assumes that the object is within the FFP region during the imaging mode and no force acts on the object.

\section{Methods and Material}
For quasi-simultaneous Magnetic Particle Imaging and Magnetic Particle Navigation (MPN), the particles need to be suitable for both applications. The final experiments are performed with an MPI scanner, but to gain insights into the properties of the particles with respect to their imaging and navigation suitability, experiments are performed in advance. For magnetic particle imaging, the particles are analyzed using a magnetic particle spectrometer. For the navigation capabilities, the separation time describing the magnetophoretic mobility is determined using a magnetic separation apparatus. 
\subsection{Magnetic Particles}
Two types of particles with different compositions and magnetic core diameters are used for the experiments. The first type is Dynabeads MyOne (ThermoFisher) with a hydrodynamic diameter of about \SI{1}{\micro\meter} as a reference. These particles consist of iron oxide embedded in a polystyrene matrix and are not suitable for \textit{in vivo} applications. The second particle type is customized nanomag/synomag-D particles (micromod Partikeltechnologie GmbH) with hydrodynamic diameters in the range of \SI{700}{\nano\meter}. These biocompatible particles consist of iron oxide cores and a dextran shell. Synomag-D particles have excellent properties as tracers for MPI,\cite{gruettner2019} but a very slow magnetic mobility. In contrast, nanomag-D particles with diameters in the range of \SIrange{250}{500}{\nano\meter} have a high magnetic mobility.\cite{henstock_remotely_2014} To combine the imaging properties of synomag-D with the high magneto-mobility of nanomag-D particles different amounts of synomag-D are embedded in the iron oxide multi-cores of nanomag and finally coated with dextran. An overview of the used particles is given in Table~\ref{tab:Particles}.
\begin{table}[hbt!]
\caption{Particles and their relevant properties used for the experiments in this work.}
\label{tab:Particles}
\centering
\begin{tabular}{ l c c c c c}	
\toprule
   Sample name &  Lot & Surface & Hydrodynamic  & Polydispersity  & Percentage of iron  \\
    & & & diameter [nm]  & index (PDI) & from synomag [\%] \\
   \midrule
  nanomag-D  &  308 & dextran  & 405  &  0.12 & -\\
  nanomag/synomag-D  &  332 &  dextran & 649 &  0.14 & 44 \\
  nanomag/synomag-D  &  333 & dextran & 698 &  0.24  & 61\\
  nanomag/synomag-D  &  334 & dextran & 641 &  0.32  & 40\\
  Dynabeads MyOne   &  017 & COOH & 1048 &  0.13 & - \\
  \bottomrule
\end{tabular}
\end{table}

\subsection{Magnetic Particle Spectrometry}
The MPS measurements are performed with a custom MPS designed similarly to the device outlined by Biederer et al.\cite{biederer2009magnetization}. The particles are excited with a sinusoidal signal with a frequency of \SI{26.042}{\kilo\hertz} and an amplitude of \SI{20}{\milli\tesla}. The measurements are performed with background subtraction and 1000 averages.

\begin{table}[hbt!]
\caption{Iron concentrations of the dilution series used for MPS measurements.}
\label{tab:serialdilution}
\centering
\begin{tabular}{ l  c c c  c ccc c c c c}
\toprule
   Num & $c_0$  & $c_1$ & $c_2$  & $c_3$  & $c_4$  & $c_5$ & $c_6$  & $c_7$ & $c_8$ & $c_9$ & $c_{10}$ \\ \midrule
   Dilution factor  &  0 &  1 &  2 & 4 &  8 &  16 & 32 & 64 &  128 & 256 &  512 \\
  Iron concentration [$\si{\milli\mol\per\liter}$]  &  179.05 &  89.52&  44.76 & 22.38 &  11.19 &  5.59& 2.79 & 1.39&  0.69&  0.35&  0.18\\
  \bottomrule
\end{tabular}
\end{table}

In order to analyze the spectra of the particles, a dilution series (summarized in Table~\ref{tab:serialdilution}) of all particle types is prepared. The analyzed volume is \SI{20}{\micro\liter} for each concentration within an Eppendorf tube. In a second measurement each particle sample at the highest concentration is magnetized with a conventional magnet. The particles remain at the bottom of the tube while the rest of the liquid becomes transparent. In a third measurement series the particles are stirred up with a vortexer. In a fourth measurement cycle the whole process of magnetization and vortexing is repeated twice. With these second, third and fourth measurements it should be ensured that the particles have no remaining remanence caused by the force mode. The particles would then have a different behavior in the imaging mode. Finally, a background measurement containing noise is subtracted from all measurements.

\subsection{Magnetic Mobility of Particles}
The magnetic mobility, in terms of its half separation time, is measured with a Q 100 ml device (SEPMAG Technologies). For these experiments \SI{100}{\milli \liter} of the particles with iron concentration of \SI{0.03}{\milli \gram \per \milli \liter} are placed inside the cavity. During the measurement the opaque suspension becomes more and more transparent until all particles have been driven towards the walls. A light source illuminates the cavity from one side to the other and the detector on the other side monitors the magnetophoresis process while the suspension homogeneity is measured over time. Afterwards, the half separation time $t_{50}$ is determined by using the Qualitance software by SEPMAG, which fits a sigmoidal curve to the measured values. A decreasing value of the half separation time represents an increase of the magnetophoretic mobility. 
In addition, the size distribution of the particle suspension before and after the magnetic separation is determined by Dynamic Light Scattering (Nano-S 90. Malvern Panalytical Ltd.).

\subsection{Magnetic Particle Navigation in Flow within Bifurcation Junction}
The particles are analyzed in an MPS system to determine their imaging capabilities and their usability for navigation has been determined in a magnetic separation device. Therefore, all subsequent experiments are performed using only the particles best suited for simultaneous imaging and navigation from the insight gained in the aforementioned experiments.

\begin{figure}[bth!]
\begin{minipage}[b]{0.3\linewidth}
  \centering
    \begin{tikzpicture}
  \node at (0,0) {\includegraphics[width=0.9\linewidth]{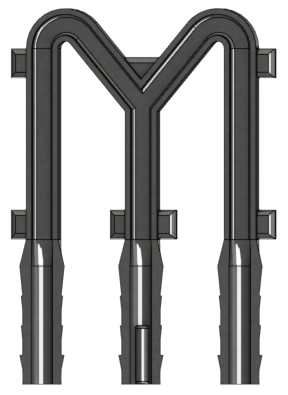}};
  \draw[->,very thick] (-2.3,-3.3)  -- (0,-3.3) node[midway,above]{y};
  \draw[->,very thick] (-2.3,-3.3)  -- (-2.3, 0) node[midway,left]{x};
  \end{tikzpicture}
  \centerline{a)}\medskip
\end{minipage}%
\begin{minipage}[b]{0.3\linewidth}
  \centering
    \begin{tikzpicture}
  \node at (0,0.0) {\includegraphics[width=1.0\linewidth]{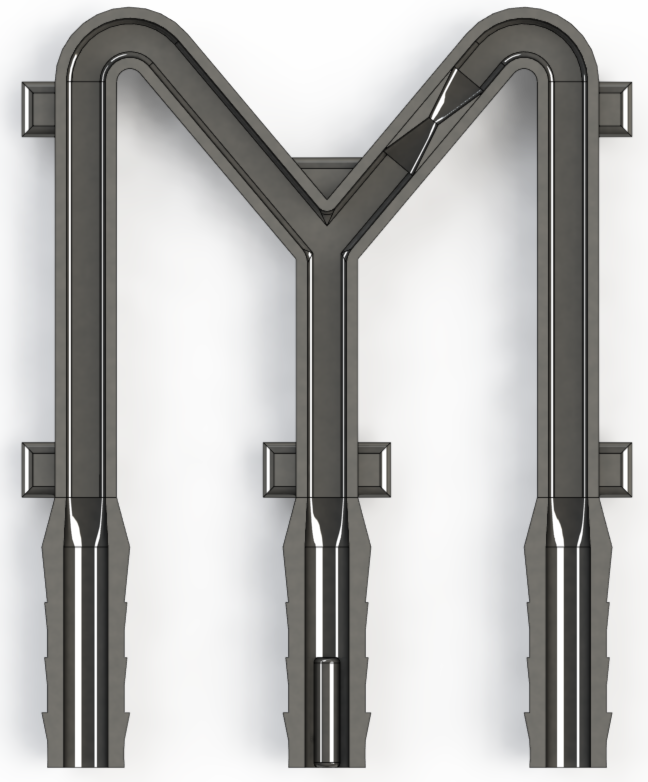}};
  \node[very thick] at (0.0,2.5) {Stenosis};
  \draw[<->,very thick] (0.7,1)  -- (0.7,-1) node[midway,right]{\SI{35}{\milli\meter}};
  \draw[<->,very thick] (0.9,1.6)  -- (1.5,2.3) node[midway,right,below,yshift=-2.5mm]{\SI{10}{\milli\meter}};
  \end{tikzpicture}
  \centerline{b)}\medskip
\end{minipage}%
\begin{minipage}[b]{0.3\linewidth}
  \centering
  \centering
    \begin{tikzpicture}
  \node at (0,0.0) {\includegraphics[width=0.9\linewidth]{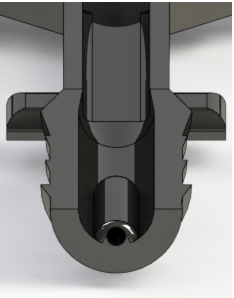}};
  \node[very thick] at (0.0,-2.7) {Centric Catheter Mount};
  \end{tikzpicture}
  \centerline{c)}\medskip
\end{minipage}
\caption{a) 3D CAD sketch of the bifurcation junction phantom. b) 3D CAD sketch of the bifurcation junction with 60\% stenosis in the right side of the phantom. c) The centric catheter mount is built centrically at the bottom of the inbound tube of every phantom to ensure centric injection of particles.}%
\label{fig:DynamicBifurcation}%
\end{figure}
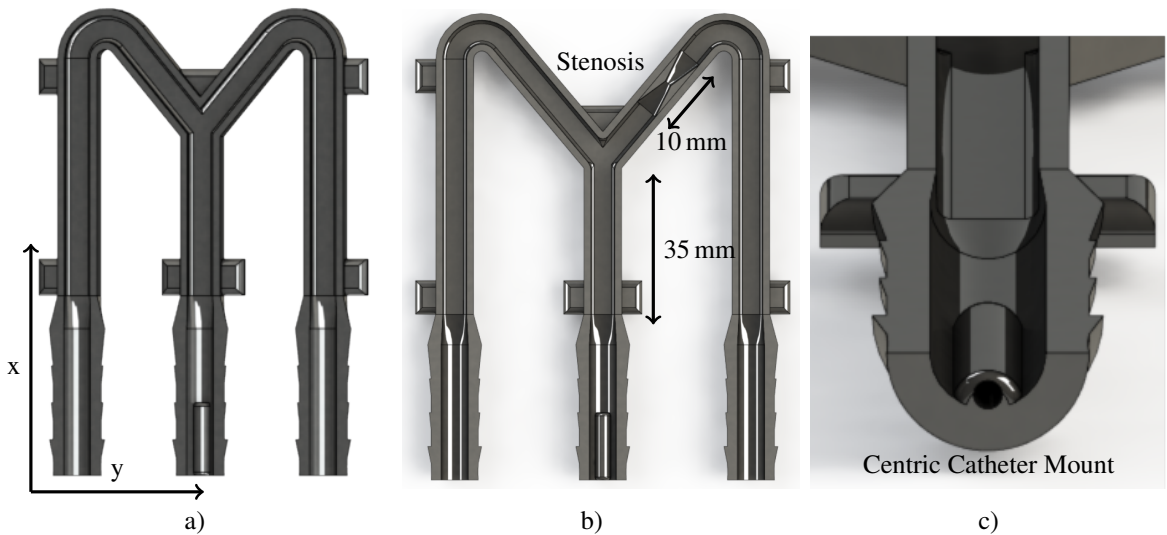

For manipulating particles in flow through a bifurcation junction using MPN, three different bifurcation phantoms are designed and 3D printed by a Forms 2 stereolithography printer from Formlabs. All bifurcation vessels have a quadratic cross-section of $A=a^2=\SI{12.544}{\square\milli\meter}$ with a side length of $a=\SI{3.544}{\milli\meter}=d_{v}$. This cross-section corresponds to the same cross-section as a circular vessel with a diameter of $d=\SI{4}{\milli\meter}$. 
The first bifurcation vessel splits into two equally-sized branches with a crossing angle of \SI{80}{\degree} as seen in Fig.~\ref{fig:DynamicBifurcation}(a). The second bifurcation phantom constitutes a 60\% stenosis in the right branch and has a circular cross-section of $A_{60}=\SI{7.52}{\milli\meter}^2$ with a diameter of \SI{1.54}{\milli\meter}. The third bifurcation phantom simulates a 100\% stenosis in the right branch and has a glued circular cross-section of $A_{100}=\SI{3.76}{\milli\meter}^2$ with a diameter of \SI{1.09}{\milli\meter} which constitutes a full blockage for the particles. Both also split with a crossing angle of \SI{80}{\degree}. The 60\% stenosis phantom is shown in Fig.~\ref{fig:DynamicBifurcation}(b). All three phantoms have a centric catheter mount in the inbound flowing branch seen in Fig.~\ref{fig:DynamicBifurcation}(c) to ensure that the particles are injected centrically.

\begin{figure}[bth!]
\begin{minipage}[b]{0.5\linewidth}
  \centering
  \centerline{\includegraphics[width=1.0\linewidth]{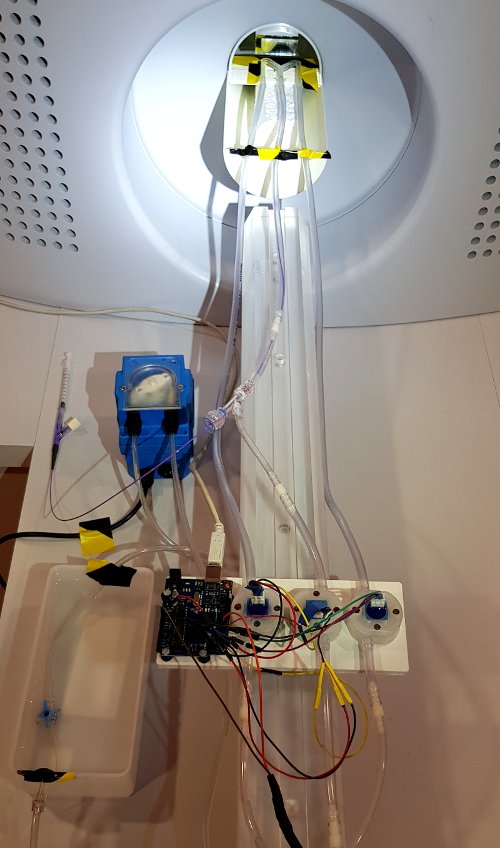}}
  \centerline{a) }\medskip
\end{minipage}%
\begin{minipage}[b]{0.5\linewidth}
  \centering
  \centerline{\includegraphics[width=1.0\linewidth]{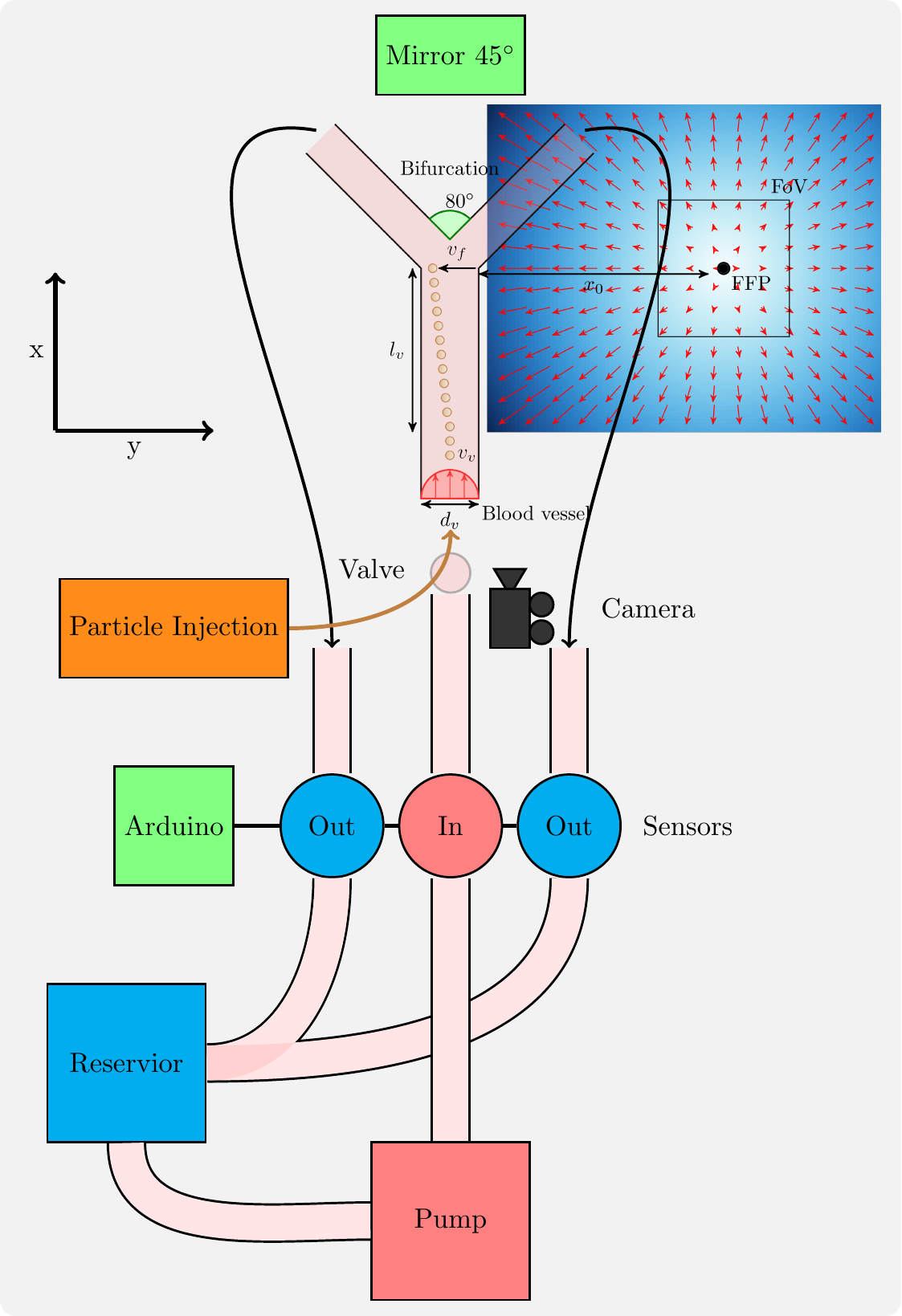}}
  \centerline{b) }\medskip
\end{minipage}
\caption{a) Bifurcation flow measurement setup with Arduino attached to flow sensors, bifurcation phantom visible inside the mirror placed inside the bore and light source. b) Schematic image of navigation of particles in bifurcation flow phantom.}
\label{fig:Exp3Real}
\end{figure}

Additionally for the dynamic flow experiments, a phantom mount is built, positioning the phantom in the center of the FoV as seen in Fig.~\ref{fig:Exp3Real}(a). A mirror at a \SI{45}{\degree} angle is fixed above the phantom to make the volume flow visible from the outside by looking through the scanner bore. A conventional light source illuminates the bore and the mirror to improve the captured optical images taken by a video camera. The bifurcation phantom is placed horizontally in the x$y$-plane to avoid the influence of the gravity force. The center of the bifurcation is placed \SI{10}{\milli\meter} off-center in the $y$-direction and \SI{5}{\milli\meter} in the $x$-direction. 
The dynamic flow bifurcation phantom is connected via three hoses to three flow sensors from BioTech with a measurement range of \SIrange{0.25}{13.33}{\milli\liter\per\second}. Both outbound hoses from the bifurcation branches are then connected back to one hose and back to the pump. The flow sensors are controlled by an Arduino which logs the flow values triggered by the MPI scanner at measurement start. The inbound hose connected to the entrance of the bifurcation phantom has a bypass enabling catheter insertion. This catheter, Abbott Armada 14 with a diameter of \SI{1.35}{\milli\meter}, is fixed in the bottom middle of the 3D-printed phantom as seen in Fig.~\ref{fig:DynamicBifurcation}(c). It ensures an unbiased particle flow at the time of the injection. The whole setup can be seen in Fig.~\ref{fig:Exp3Real}(a) and Fig.~\ref{fig:Exp3Real}(b).

The preclinical MPI scanner used generates a maximum gradient of \SI{2.4}{\tesla\per\meter} in the $z$-direction (\SI{1.2}{\tesla\per\meter} in the $x$- and $y$-directions). The FFP offset can be set to a maximum of $\SI{-42}{\milli\tesla}$ to $\SI{42}{\milli\tesla}$ ($\SI{-17.5}{\milli\meter}$ to $\SI{17.5}{\milli\meter}$) in the $z$-direction and $\SI{-17}{\milli\tesla}$ to $\SI{17}{\milli\tesla}$ ($\SI{-14.1}{\milli\meter}$ to $\SI{14.1}{\milli\meter}$) in the $x$- and $y$-directions. This leads to a maximum force on the particles of $1.18\cdot 10^{-12}$~N in the $z$-direction and $2.96\cdot 10^{-13}$~N in the $y$-direction, by assuming a particle diameter of \SI{1}{\micro\meter}. 
In our case the FFP in force mode is adjusted \SI{14.16}{\milli\meter} away in the opposed $y$-direction, making a total distance from bifurcation center to the FFP of \SI{22.16}{\milli\meter} ($4.64 \cdot 10^{-13}$~N). In the $x$-direction the FFP is also set \SI{14.16}{\milli\meter} away from the MPI FoV center, making a total distance between FFP and bifurcation center of \SI{9.16}{\milli\meter}. 
For these experiments we only use the MPN (DF-cycles $\varphi=20$) which means no imaging (DF-cycles $\zeta=0$) is done and the force acts all the time on the particles.

The following flow experiments with the bifurcation phantom are conducted with five different flow rates from \SIrange{5.45}{10.22}{\milli\liter\per\second} in the inflowing branch. For each flow velocity the experiment is performed twice. First, the magnetic forces are turned off to gain ground truth flow behavior and exclude any biased flow direction. This ensures that the particles flow equally through both branches. Secondly, in the actual experiment the magnetic forces are turned on to investigate the particles' flow characteristics influenced by the magnetic field. To eliminate a bias to one side of the bifurcation branch and to show that the method works in both directions of the bifurcation, the experiment is conducted for both branches of the bifurcation for the flow rate \SI{5.45}{\milli\liter\per\second}.

\begin{table}[h!bt]
\caption{Volumetric flow rates used in the measurements and the length of time the particles are influenced by the magnetic force.}
\label{tab:Sensorflowrate}
\centering
\begin{tabular}{ l ccc }
\toprule
   Degree of stenosis & Flow rate  & Flow rate   & Time [ms]\\
     &inbound [\si{\milli \liter \per \second}]&outbound [\si{\milli \liter \per \second}]&\\ \midrule
   100\% & 1.36 & 0.68 & 322 \\
   0\%, 60\%, 100\% &  2.72& 1.36 & 161\\
   0\%, 60\% &  5.45 & 2.72 & 80 \\
  0\%, 60\% &  6.87 & 3.40 & 63 \\
  0\%, 60\% &  8.18 & 4.09 & 53\\
  0\%  &  10.22 & 5.11 & 42 \\
  \bottomrule
\end{tabular}
\end{table}
The measured flow values from the sensors coincide with the theory. The flow in both branches is the same and their sum corresponds to the flow values inbound to the bifurcation. The five investigated flow velocities are given in Table~\ref{tab:Sensorflowrate}. The distance from the location where particles are released to the center of the bifurcation is $l_v$ = \SI{35}{\milli\meter}. The particles are ejected with a catheter fixed centrically inside the vessel phantom. The time span from particle release point to the center of the bifurcation is also given in the final column of Table~\ref{tab:Sensorflowrate}.

\subsection{Magnetic Particle Imaging}
MPI with simultaneous navigation application is performed using the preclinical MPI scanner, which is also used in the navigation only experiments. Imaging of static particles without navigation application is performed in advance to check the imaging capabilities of the particles in the MPI scanner. 
Two system matrices are acquired on a grid of \num{24x24x14} positions covering a volume of \SI{24x24x14}{\milli \meter} and delta samples of \SI{1x 1x 1}{mm} filled with \SI{1}{\micro \liter} of Dynabeads MyOne and nanomag/synomag-D 333. The gradient strength is set to \SI{2.4}{\tesla\per\meter} and the drive-field amplitude to \SI{12}{\milli \tesla} resulting in a FoV of \SI{20x20x10}{\milli \meter}. For the imaging measurement the delta samples of \SI{1x 1x 1}{mm} are used respectively. The imaging sequence is performed with the same imaging parameters to enable reconstruction using a regularized iterative Kaczmarz algorithm. Before reconstruction, the data from 100 DF-cycles are averaged and frequencies with an SNR above two are selected from the frequency range of \SIrange{80}{1250}{\kilo \hertz}, resulting in 423 frequency components. The relative regularization parameter is set to 0.001 and three iterations are performed. Afterwards, the full width at half maximum (FWHM) is determined in the $x$- and $z$-directions within the image.

\subsection{Simultaneous Magnetic Particle Imaging and Navigation of Phantom with Flow in Bifurcation Junction}
The simultaneous MPIN bifurcation experiments within flow are only performed for flow velocity \SI{1.36}{\milli\liter\per\second} and only with the bifurcation phantom with 100\% stenosis. The position of the imaging FoV is adjusted by FF$_{\text{I,xyz}}$=\SI{0x14.1x0}{\milli\meter} with the help of the focus fields to cover the phantom branch with the simulated stenosis. Due to field deviations the system matrix acquired for the previous imaging experiments cannot be used for the imaging part of the experiments with simultaneous navigation application. Thus, a system matrix with the same imaging parameters used in the previous experiments is acquired at the shifted focus field position. The focus field position for navigation is set to FF$_{\text{F,xyz}}$=\SI{14.1x-14.1x0}{\milli \meter}.
For simultaneous imaging and navigation the number of imaging DF-cycles is set to $\zeta=1$ while the number of navigation DF-cycles is set to $\varphi=20$. The FFP then switches between FF$_{\text{I,xyz}}$ and FF$_{\text{F,xyz}}$ with the ratio of 20:1 taking snapshots at the 100\% stenosis. Overall, this results in a temporal resolution for imaging of \SI{2.15}{\second} per image.

\section{Results}

\subsection{Magnetic Particle Spectrometer}
\begin{figure}[htb!]%
\begin{minipage}[b]{0.33\linewidth}
  \centering
  \centerline{\includegraphics[width=1.0\linewidth]{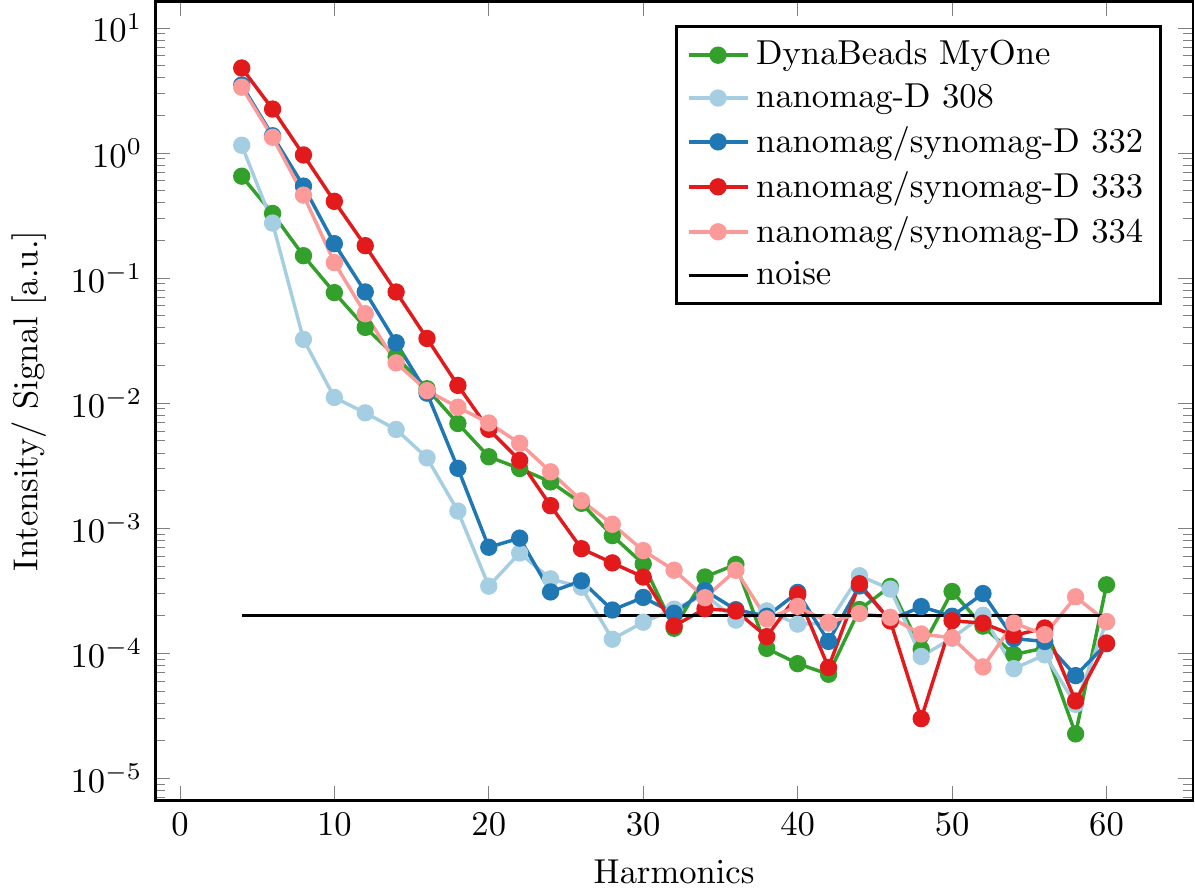}}
  \centerline{(a) Spectra }\medskip
\end{minipage}
\begin{minipage}[b]{0.33\linewidth}
  \centering
  \centerline{\includegraphics[width=1.0\linewidth]{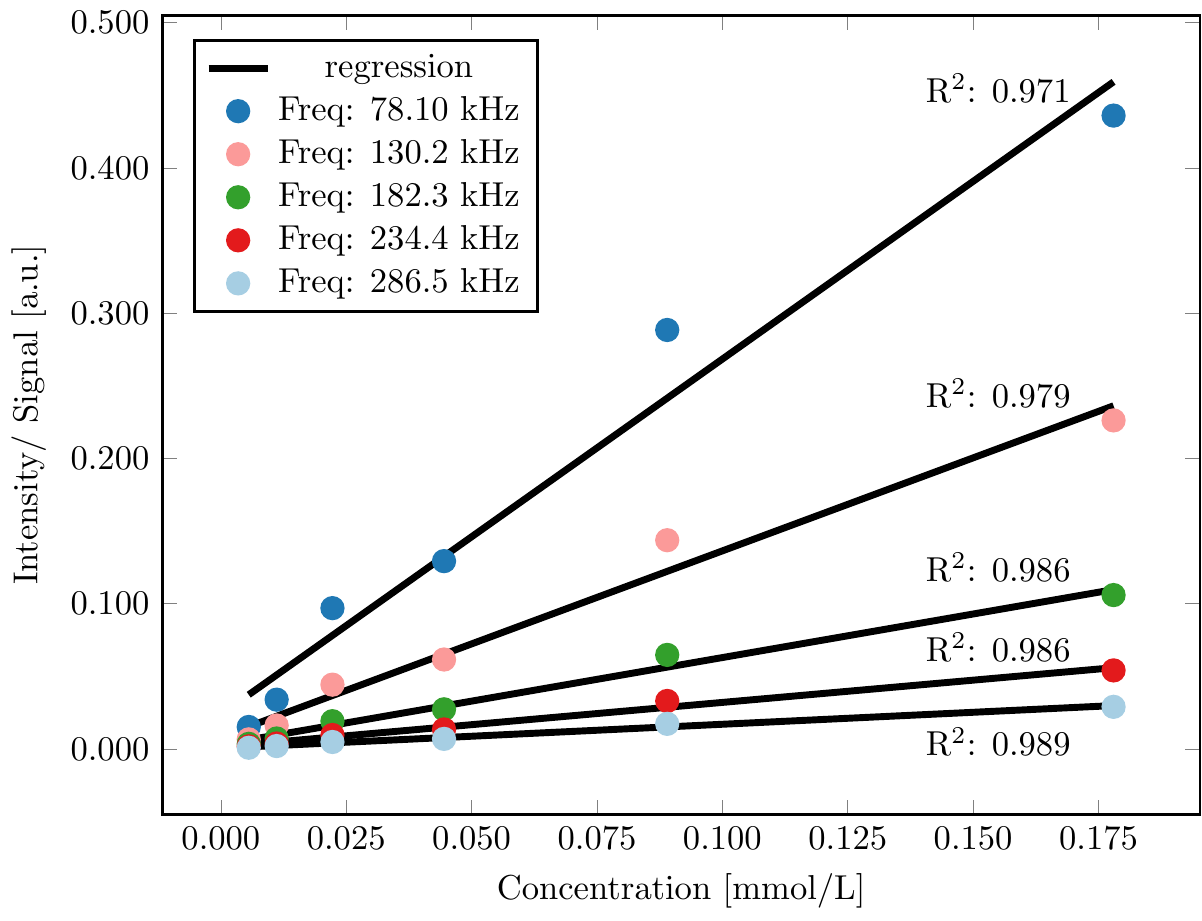}}
  \centerline{(b) Dynabeads MyOne}\medskip
\end{minipage}
\begin{minipage}[b]{0.33\linewidth}
  \centering
  \centerline{\includegraphics[width=1.0\linewidth]{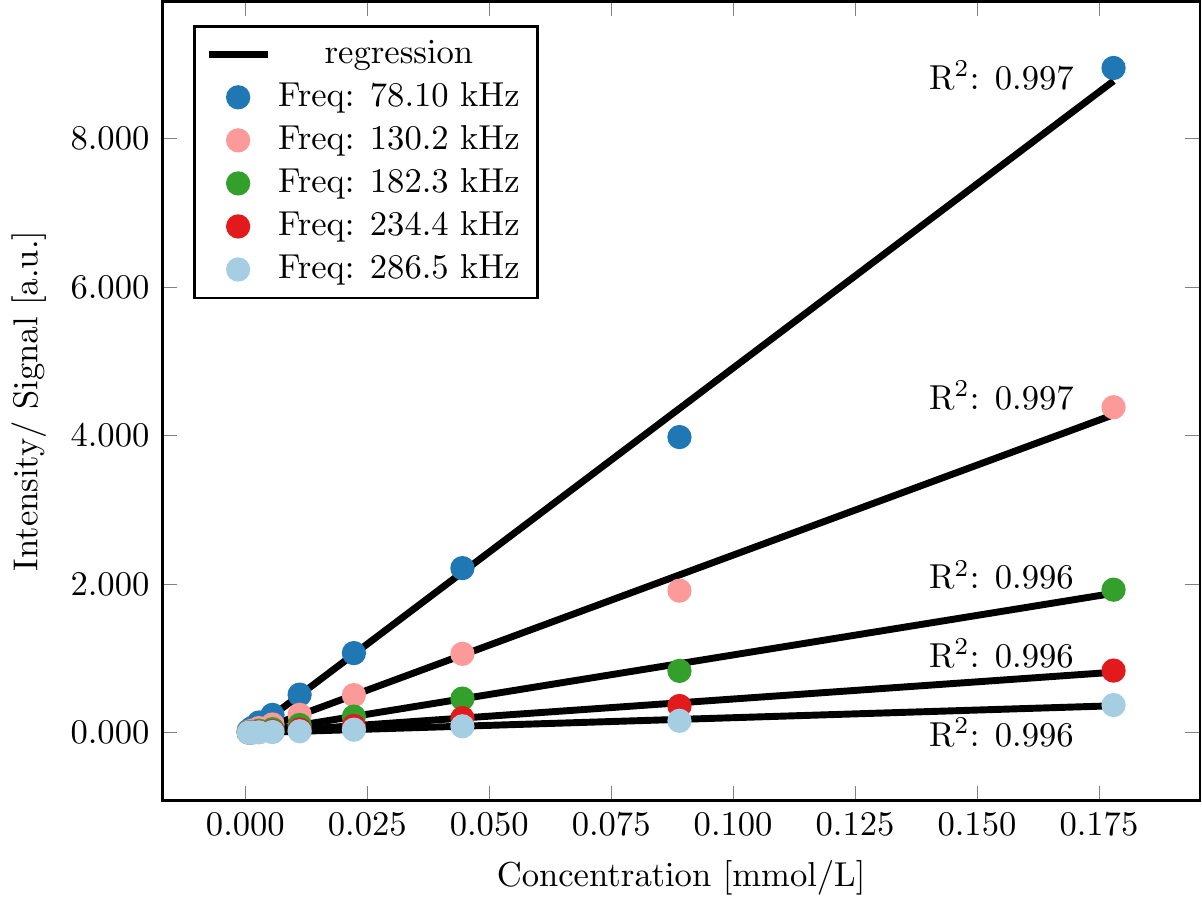}}
  \centerline{(c) nanomag/synomag-D 333.}\medskip
\end{minipage}
\caption{(a) Spectra of Dynabeads MyOne, nanomag/synomag-D batches 332,333,334,308.  (b) Linear signal strength decrease for various frequencies dependent on the concentration of beads. (c) Linear signal strength decline for various frequencies dependent on the concentration of nanomag/synomag-D batch 333.}
\label{fig:TimeLinearity}%
\end{figure}

In Fig.~\ref{fig:TimeLinearity}(a) the spectra of Dynabeads MyOne and four different nanomag/synomag-D particle types in a concentration of \SI{179.05}{\milli\mol\per\liter} are shown. The batches (332, 308) only generate up to 28 harmonics and the signal decrease is not linear. The spectra of the Dynabeads MyOne, and the nanomag/synomag-D batches (333, 334) generate up to 35 harmonics above noise level. Batch 333 has the strongest signal of all batches but also the steepest linear constant decline of the signal strength as seen in Fig.~\ref{fig:TimeLinearity}(a). Batch 334 has lower signal in the beginning harmonics but its steepness is lower and has a kink at the 18$^{th}$ harmonic where the curvature of the signal strength becomes even less steep. For further investigations, only the results of nanomag/synomag-D batch 333 are shown because batch 333 provides the best result of all batches in the spectrum and is therefore the most promising candidate for imaging abilities.

In Fig.~\ref{fig:TimeLinearity}(b) and Fig.~\ref{fig:TimeLinearity}(c) the linearity within a frequency over different concentrations is investigated for Dynabeads MyOne and nanomag/synomag-D batch 333. It shows that the signal strength in all frequencies is linear, with coefficients of determination between 0.971 and 0.997.

\begin{figure}[htb!]%
\begin{minipage}[b]{0.33\linewidth}
  \centering
  \centerline{\includegraphics[width=1.0\linewidth]{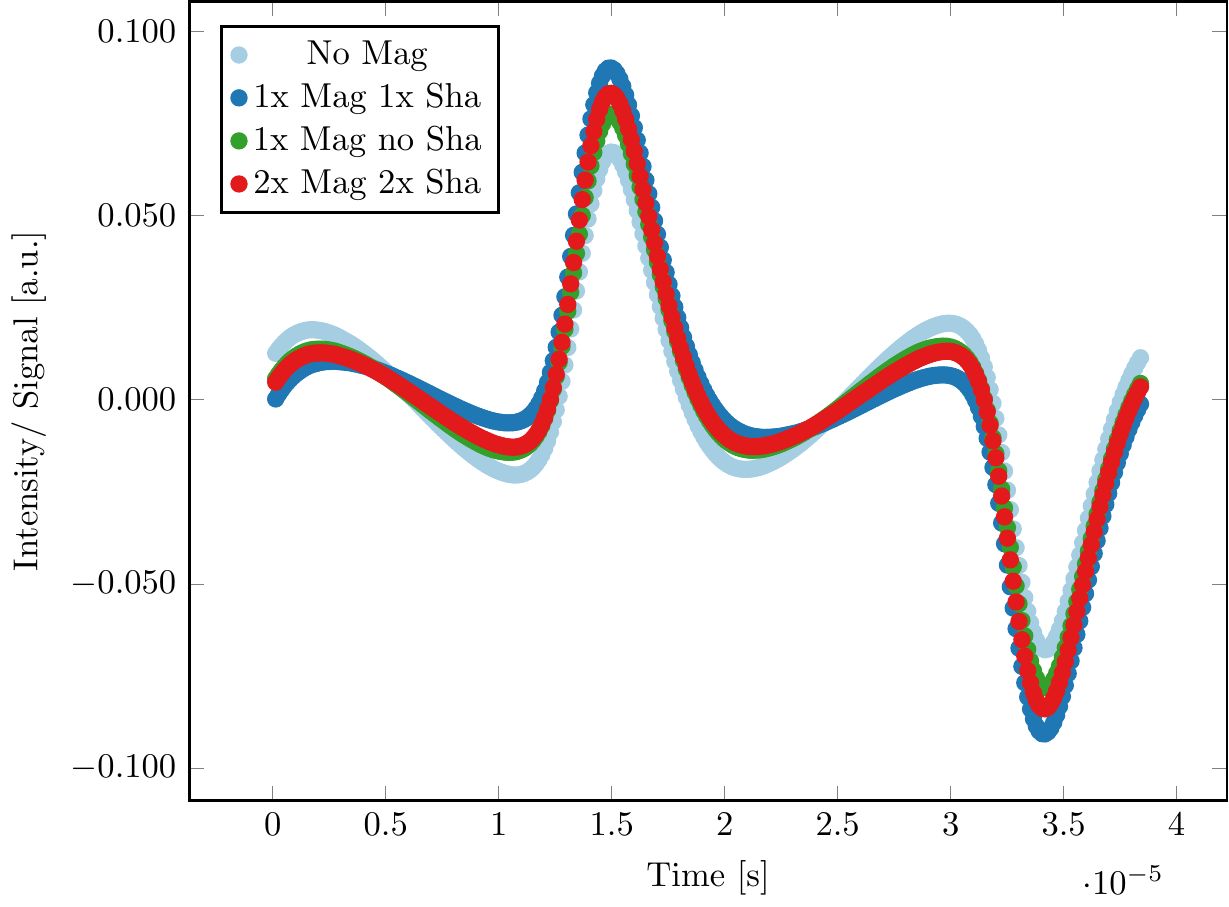}}
  \centerline{(a) Dynabeads MyOne}\medskip
\end{minipage}
\begin{minipage}[b]{0.33\linewidth}
  \centering
  \centerline{\includegraphics[width=1.0\linewidth]{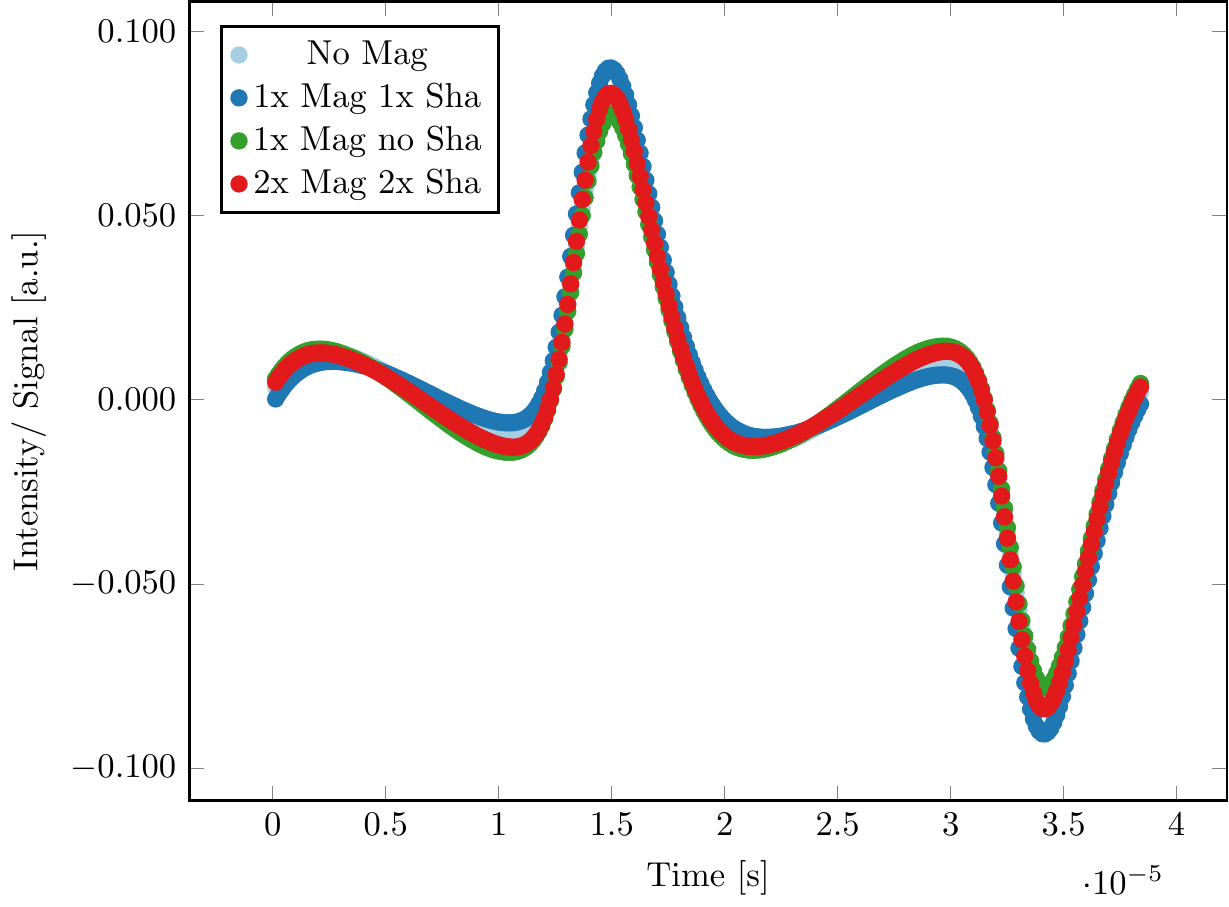}}
  \centerline{(b) nanomag/synomag-D 333.}\medskip
\end{minipage}
\caption{(a) Time signal for Dynabeads MyOne in four different stages: non-magnetized, magnetized, magnetized and vertex, 2x magnetized and 2x vertex. (b) Time signal for nanomag/synomag-D 333 in four different stages: non-magnetized, magnetized, magnetized and vertex, 2x magnetized and 2x vertex. }
\label{fig:Magnetization}%
\end{figure}

Since the particles should be used for magnetic navigation applications, their imaging characteristics are investigated for any effects of magnetization of the particles. In Fig.~\ref{fig:Magnetization}(a) the time signal of beads at concentration \SI{179.05}{\milli\mol\per\liter} is shown for different stages. The light blue line presents the signal of non-magnetized particles, while in the dark blue line the particles have been magnetized and settled at the bottom of the tube. The results in the green line are from particles that have been magnetized and vertexed once. The results in the red line are from particles that went through this procedure twice. All four time signals seem similar with only very small deviations.
The same four lines are presented in Fig.~\ref{fig:Magnetization}(b) for the nanomag/synomag-D batch 333 particles. Here, all graphs also appear very similar with only very small discrepancies. The magnetization attempt and vertexing do not seem to influence the particles' imaging behavior.

\subsection{Magnetic Mobility of Particles}
\begin{figure}[hbt]
  \centering
    \includegraphics[width=0.5\linewidth]{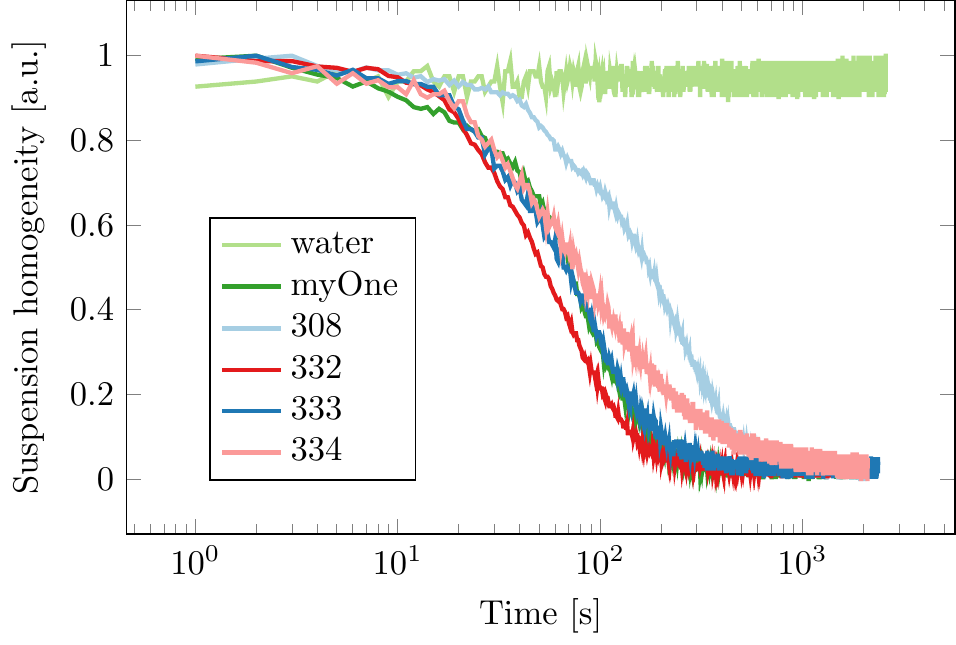}
    \caption{During the magnetic separation the homogeneity of the suspension is measured over time. The suspension homogeneity is plotted for all particle batches, while water is used as a blank value for comparison.}
    \label{fig:MagnetophoreticCurve}
\end{figure}

The suspension homogeneity during the separation process for the different particles is shown in Fig.~\ref{fig:MagnetophoreticCurve}. The homogeneity has been normalized with the initial homogeneity. The relevant navigation parameters calculated based on these time curves (the half separation time values), as well as the size distribution of the particle suspension before and after the magnetic separation, and the polydispersity index (PdI) are given in Table~\ref{tab:MagnetophoreticResults}. 
The half separation times (\SI{75}{\second}, \SI{83}{\second}) of nanomag/synomag-D batches (333, 334) are within the same order as the half separation of the Dynabeads MyOne at \SI{79}{\second}. In contrast, the half separation time of nanomag/synomag-D (332) particles is much shorter at \SI{57}{\second}, whereas the half separation time of plain nanomag-D is much longer at \SI{194.4}{\second}.
\begin{table}[hbt!]
\caption{Half-separation times $t_{50}$ of the different suspensions of iron oxide particles with water as a blank value and size distribution before and after separation. }
\label{tab:MagnetophoreticResults}
\centering
\begin{tabular}{ l cc p{2.5cm} p{2.5cm}}	
\toprule
   Sample name & Lot  & $t_{50}$[s] & $Z_\text{average}$[nm] / PdI before separation & $Z_\text{average}$[nm] / PdI after separation \\ \midrule
  nanomag-D  &  308 &  194.4 &  405.4 / 0.118 & 505.0 / 0.229   \\
  nanomag/synomag-D  &  332 &  57.0 &  649.2 / 0.143  & 679.2 / 0.171 \\
  nanomag/synomag-D  &  333 &  75.0 &  698.2 / 0.240  & 1104.0 / 0.297 \\
  nanomag/synomag-D  &  334 &  83.0 &  641.3 / 0.325  & 770.2 / 0.481 \\
  Dynabeads MyOne COOH  &  017 &  79.0 &  1048 / 0.126  & 1105.0 / 0.280 \\
  \bottomrule
\end{tabular}

\end{table}
The size distribution is influenced by the separation process resulting in an increase of $Z_{\text{average}}$. The increase is \SI{130}{\nano \meter} at its maximum after the magnetic separation for all particles except for nanomag/synomag-D 333 particles. For these particles the difference is \SI{405}{\nano \meter} before and after the magnetic separation process.

\subsection{Navigation Experiments in Flow}
The results are shown in videos. An overview of the flow measurement results can be found in Table~\ref{tab:Exp3} with video reference, flow parameters for inbound and outbound bifurcation branches, left or right bifurcation and description.

\begin{threeparttable}[hbt!]
\caption{Video results for bifurcation flow measurements and their controls for different flow velocities.}
\label{tab:Exp3}
\centering
\begin{tabular}{ l  c  c  p{10cm}  }		
\toprule
   Video ref & Flow rate inbound / &  Bifurcation & Description of results \\
   &outbound [ml/s]&side&\\ \midrule
  v1.1.0R & 2.72 / 1.36 &   right & Particles flow only through the right branch   \\ 
  v1.1.1R & 2.72 / 1.36 &   right & Control No Force: Particles flow through both branches equally  \\
  v1.1.0L & 2.72 / 1.36 & left   & Particles flow only through the left branch  \\
  v1.1.1L & 2.72 / 1.36 & left   & Control No Force: Particles flow through both branches equally  \\
  v1.2.0L & 5.45 / 2.72 & left  & Particles flow only through the left branch   \\
  v.1.2.1L\tnote{*}  & 5.45 / 2.72 & left  &   Control No Force: Particles flow through both branches equally  \\
  v1.3.0L & 6.87 / 3.40 & left   & Particles flow only through the left branch  \\
  v1.3.1L & 6.87 / 3.40 & left   & Control No Force: Particles flow through both branches equally   \\
  v1.4.0L & 8.18 / 4.09 & left  & Particles flow mainly through the left branch, depending on injection pressure  \\
  v1.4.1L & 8.18 / 4.09 & left  & Control No Force: Particles flow through both branches equally  \\
  v1.5.0L & 10.22 / 5.11 & left  & Effect still visible but, due to the high flow velocity, not all particles are navigated to the left.  \\
  \bottomrule
\end{tabular}
\begin{tablenotes}\footnotesize 
\item[*] video broken
\end{tablenotes}
\end{threeparttable}
\vspace{1em}

In general, it can be observed that magnetic forces from the FFP steer the particles to one side of the bifurcation for flow velocities up to \SI{8.18}{\milli\liter\per\second} in the inbound branch. The method works for both sides of the bifurcation depending on the position of the FFP. The control measurement shows that the particles flow equally through both branches and a bias for one direction is negligibly low. At the flow velocity \SI{10.22}{\milli\liter\per\second} inbound, the majority of particles  steer to the left branch, while the remaining particles travel too quickly through the navigation window and flow with the current in the right branch.
\def\w1{-2.15} 
\def\h1{-1.70} 
\def\lwp{0.24}
\begin{figure}[htb!]%
\begin{minipage}[b]{\lwp\linewidth}
  \centering
  \begin{tikzpicture}
  \node at (0,0) {\includegraphics[width=1.0\linewidth]{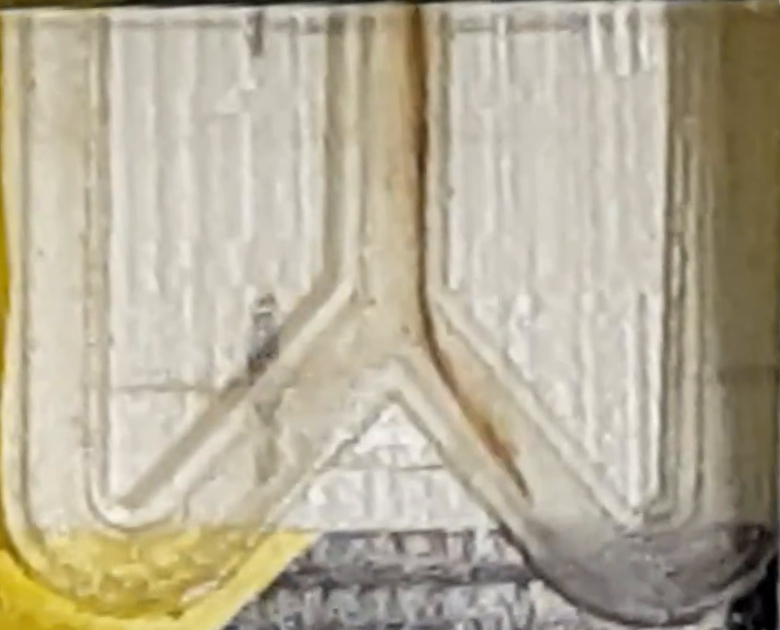}};
  \draw[->,ultra thick] (-1.5,1) -- (-0.5,1) node[midway,below]{Force $F_y$};
  \draw[->,ultra thick] (\w1,\h1)  -- (\w1,0) node[midway,left]{x};
  \draw[->,ultra thick] (\w1,\h1)  -- (0.0,\h1) node[midway,below]{y};
  \draw[->,very thick] (0.1,-1.6)   -- node[midway,right]{$F_x$} (0.1,-0.9);
  \end{tikzpicture}
  \centerline{\parbox{4.0cm}{(a) $F_y>0$, no stenosis, flow rate \SI{2.72}{\milli\liter\per\second} }}\medskip
\end{minipage}%
\begin{minipage}[b]{\lwp\linewidth}
  \centering
  \begin{tikzpicture}
  \node at (0,0) {\includegraphics[width=1.0\linewidth]{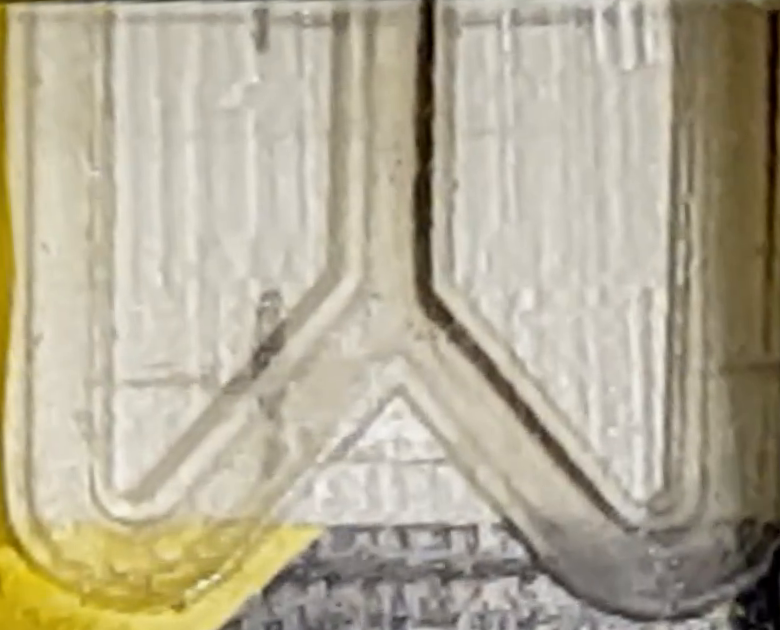}};
  \draw[->,ultra thick] (-1.5,1) -- (-0.5,1) node[midway,below]{Force $F_y$};
  \draw[->,ultra thick,opacity=0.0] (\w1,\h1)  -- (\w1,0) node[midway,left]{x};
  \draw[->,ultra thick,opacity=0.0] (\w1,\h1)  -- (0.0,\h1) node[midway,below]{y};
  \draw[->,very thick] (0.1,-1.6)   -- node[midway,right]{$F_x$} (0.1,-0.9);
  \end{tikzpicture}
  \centerline{\parbox{4.0cm}{(b) $F_y>0$, no stenosis, flow rate \SI{2.72}{\milli\liter\per\second}}}\medskip
\end{minipage}%
\begin{minipage}[b]{\lwp\linewidth}
  \centering
    \begin{tikzpicture}
  \node at (0,-0.075) {\includegraphics[width=1.0\linewidth]{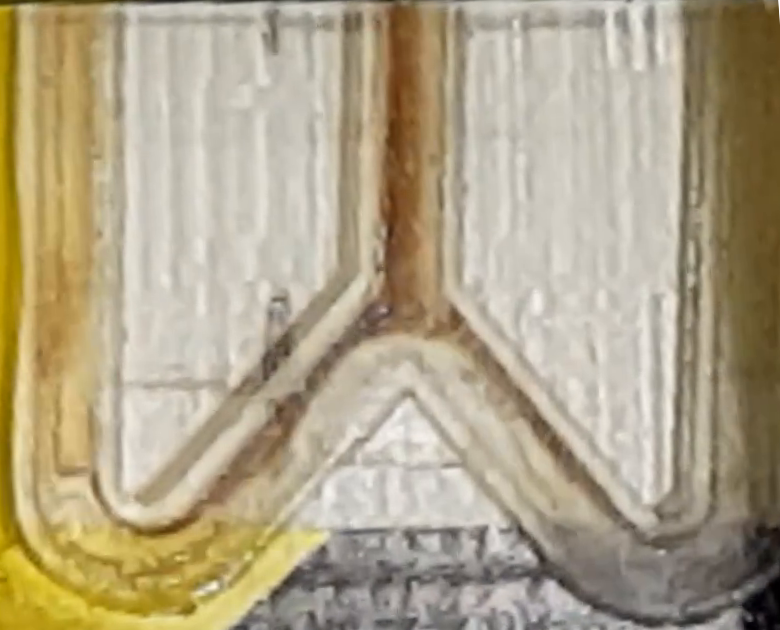}};
  \node[very thick] at (-1,1) {Force $F_y$=0};
  \draw[->,ultra thick,opacity=0.0] (\w1,\h1)  -- (\w1,0) node[midway,left]{x};
  \draw[->,ultra thick,opacity=0.0] (\w1,\h1)  -- (0.0,\h1) node[midway,below]{y};
  \end{tikzpicture}
  \centerline{\parbox{4.0cm}{(c) Control: $F_y=0$, No stenosis, flow rate \SI{2.72}{\milli\liter\per\second} }}\medskip
\end{minipage}
\caption{a) First time point in the flow measurement with magnetic forces and flow velocity \SI{2.72}{\milli \liter \per \second} where particles are injected and flow through the right side of the bifurcation. (b) Second time point where particles are injected and pushed to the right side and remain motionless at the side, as long as the magnetic fields are turned on. c) Control measurement where forces are turned off and particles flow equally through both bifurcation branches.}
\label{fig:ResultExp3Time1_2}%
\end{figure}
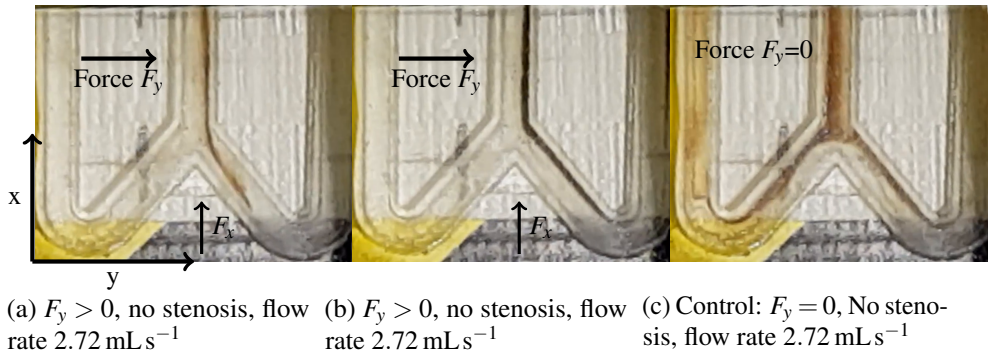

In Fig.~\ref{fig:ResultExp3Time1_2}(a) and Fig.~\ref{fig:ResultExp3Time1_2}(b) examples of two time points during the injection of particles are shown for the flow velocity \SI{2.72}{\milli \liter \per \second} inbound. The particles are always injected centrically in the phantom and the particles are pushed to the right side and enter the right branch as seen in Fig.~\ref{fig:ResultExp3Time1_2}(a). At the second time point the particles also move to the right side and enter the right branch but they remain on the right side despite the flow pushing them forward. These particles remain motionless on the right side, as depicted in Fig.~\ref{fig:ResultExp3Time1_2}(b), as long as the magnetic field sequence is active. After the magnetic fields are turned off the particles are released by the outbound flow. In the control measurement with force $F_y=0$, seen in Fig.~\ref{fig:ResultExp3Time1_2}(c), the particles are interfused with water and they are distributed equally towards both sides of the bifurcation.
\def\w2{-2.15} 
\def\h2{-1.65} 
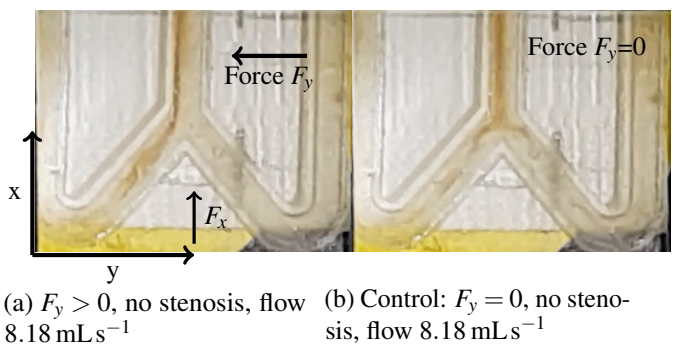
\begin{figure}[htb!]%
\begin{minipage}[b]{\lwp\linewidth}
  \centering
  \begin{tikzpicture}
  \node at (0,0) {\includegraphics[width=1.0\linewidth]{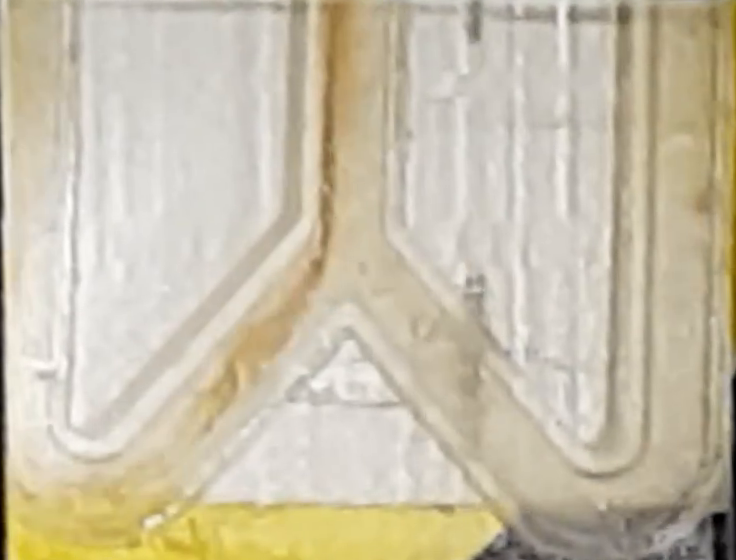}};
  \draw[->,ultra thick] (1.5,1) -- (0.5,1) node[midway,below]{Force $F_y$};
  \draw[->,ultra thick] (\w2,\h2)  -- (\w2,0) node[midway,left]{x};
  \draw[->,ultra thick] (\w2,\h2)  -- (0.0,\h2) node[midway,below]{y};
  \draw[->,very thick] (0.0,-1.5)   -- node[midway,right]{$F_x$} (0.0,-0.8);
  \end{tikzpicture}
  \centerline{\parbox{4.0cm}{(a) $F_y>0$, no stenosis, flow \SI{8.18}{\milli\liter\per\second}} }\medskip
\end{minipage}%
\begin{minipage}[b]{\lwp\linewidth}
  \centering
  \begin{tikzpicture}
  \node at (0,-0.075) {\includegraphics[width=1.0\linewidth]{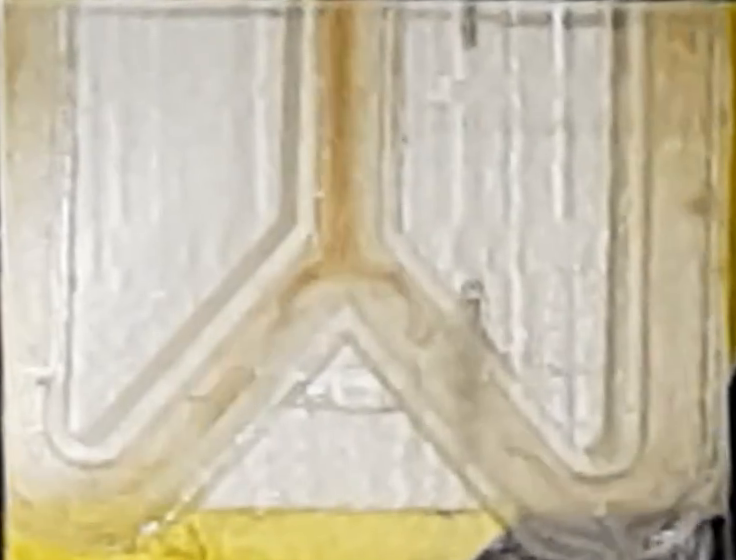}};
  \node[very thick] at (1,1) {Force $F_y$=0};
  \draw[->,ultra thick,opacity=0.0] (\w2,\h2)  -- (\w2,0) node[midway,left]{x};
  \draw[->,ultra thick,opacity=0.0] (\w2,\h2)  -- (0.0,\h2) node[midway,below]{y};
  \end{tikzpicture}
  \centerline{\parbox{4.0cm}{(b) Control: $F_y=0$, no stenosis,  flow \SI{8.18}{\milli\liter\per\second} }}\medskip
\end{minipage}
\caption{(a) The particles are entirely pushed to the left side of the bifurcation at a flow rate of \SI{8.18}{\milli\liter\per\second}. (b) When the magnetic fields are turned off in the control measurement the particles distribute equally to both branches at a flow velocity of \SI{8.18}{\milli\liter\per\second}.}
\label{fig:ResultExp3Time3_4}%
\end{figure}

The highest flow velocity in our experiments with successful navigation of particles towards one of the bifurcation branches is determined to be \SI{8.18}{\milli\liter\per\second}. This is seen in Fig.~\ref{fig:ResultExp3Time3_4}(a) for a tube phantom with a cross-section of $A=\SI{12.544}{\milli\meter}^2$ and an inbound length of $l=\SI{35}{\milli\meter}$. In the control measurement shown in Fig.~\ref{fig:ResultExp3Time3_4}(b) with no force acting on the particles, the particles are distributed equally to both sides of the bifurcation.
The particles can only be stopped at the side of the inbound tube at the velocity \SI{2.72}{\milli\liter\per\second}. In contrast, this effect does not occur at higher flow velocities.

\subsubsection{Flow analysis with bifurcation and 60\%-100\% stenosis}
Further flow experiments are conducted using a bifurcation phantom with a 60\% stenosis in the right branch. The experiments are performed using different flow velocities.
For all velocities up to \SI{6.87}{\milli\liter\per\second} it is possible to navigate the particles to the right branch. However, all particles flow through the left branch during the control measurement, where the high flow velocity in the left side is caused by the 60\% stenosis. Only at velocity \SI{8.18}{\milli\liter\per\second}, the particles flow through both, the left and right branch. The video references of the 60\% stenosis measurements and their control measurements for different velocities can be found in Table~\ref{tab:Exp3Stenosis}.

\begin{table}[hbt!]
\caption{Video results for 60\% stenosis measurements and their controls for different flow velocities.}
\label{tab:Exp3Stenosis}
\centering
\begin{tabular}{ c c c p{9cm}  }	
\toprule
   Video ref & Flow rate inbound/ &  Bifurcation & Description of results  \\
   &outbound [ml/s]&&\\ \midrule
  v2.1.0R & 2.72  / 1.36   & right & Particles flow only through stenosis in right branch \\
  v2.1.1R & 2.72  / 1.36  & right & Control: Particles flow mainly through left branch  \\
  v2.2.0R & 5.45 / 2.72  & right & Particles flow only through stenosis in right branch  \\
  v2.2.1R & 5.45 / 2.72  & right & Control: Particles flow mainly through left branch  \\
  v2.3.0R & 6.87 / 3.40   & right & Particles flow mostly through stenosis in right branch  \\
  v2.3.1R & 6.87 / 3.40   & right & Control: Particles flow mainly through left branch    \\
  v2.4.0R & 8.18 / 4.09  & right & Particles flow through left branch and stenosis in right branch \\
  v2.4.1R & 8.18 / 4.09   &  right & Control: Particles flow mainly through left branch  \\
  \bottomrule
\end{tabular}
\end{table}

\def\w3{-2.15} 
\def\h3{-1.70} 
\begin{figure}[htb!]%
\begin{minipage}[b]{\lwp\linewidth}
  \centering
  \begin{tikzpicture}
  \node at (0,0) {\includegraphics[width=1.0\linewidth]{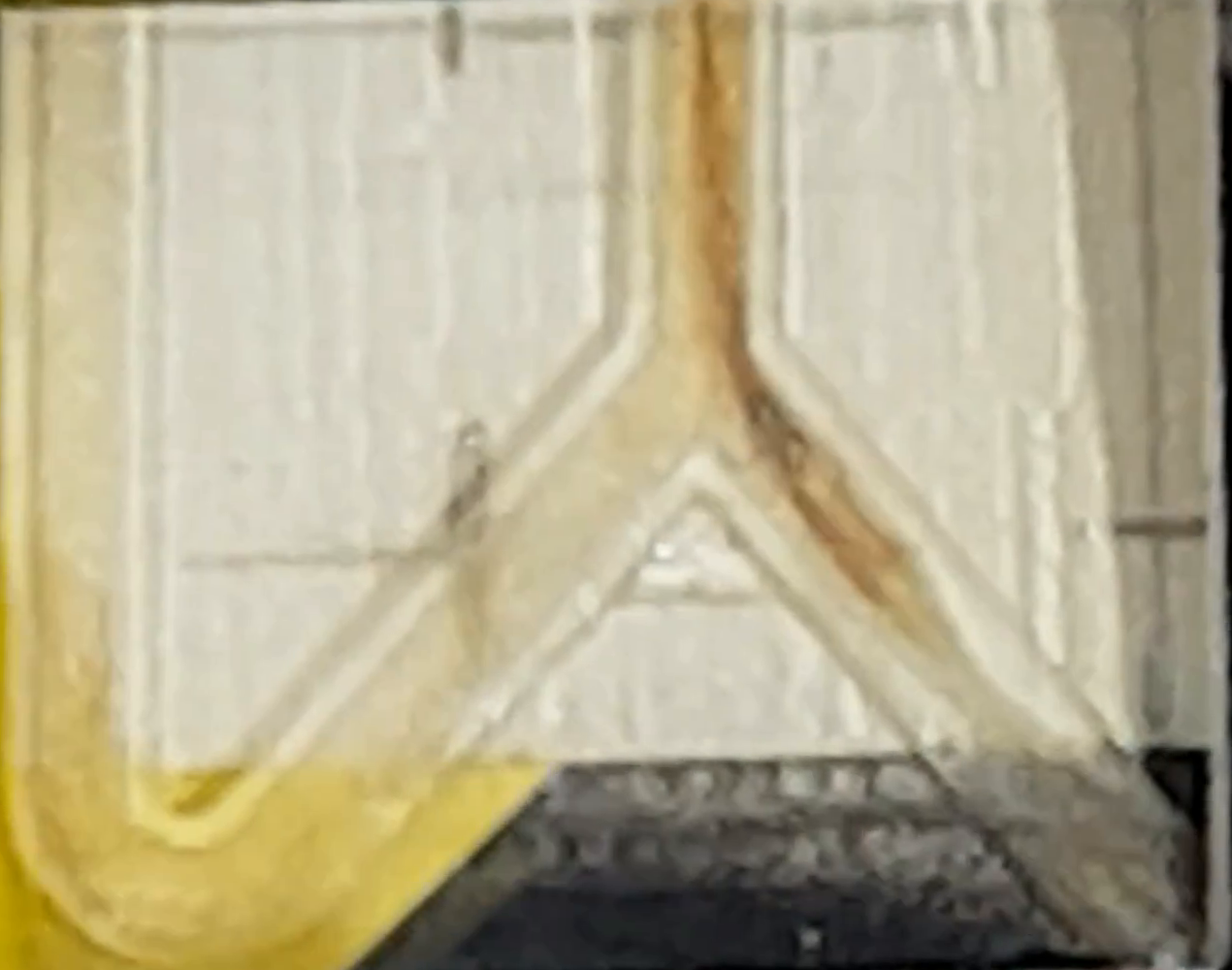}};
  \draw[->,ultra thick] (-1.5,1) -- (-0.5,1) node[midway,below]{Force $F_y$};
  \draw[->,very thick] (1.5,0.8) node[above]{Stenosis} -- (1.4,-0.2);
   \draw[->,ultra thick] (\w3,\h3)  -- (\w3,0) node[midway,left]{x};
  \draw[->,ultra thick] (\w3,\h3)  -- (0.0,\h3) node[midway,below]{y};
  \end{tikzpicture}
  \centerline{\parbox{3.8cm}{(a) $F_y>0$, 60\% stenosis, velo \SI{6.87}{\milli\liter\per\second}} }\medskip
\end{minipage}
\begin{minipage}[b]{\lwp\linewidth}
  \begin{tikzpicture}
  \node at (0,-0.075) {\includegraphics[width=1.0\linewidth]{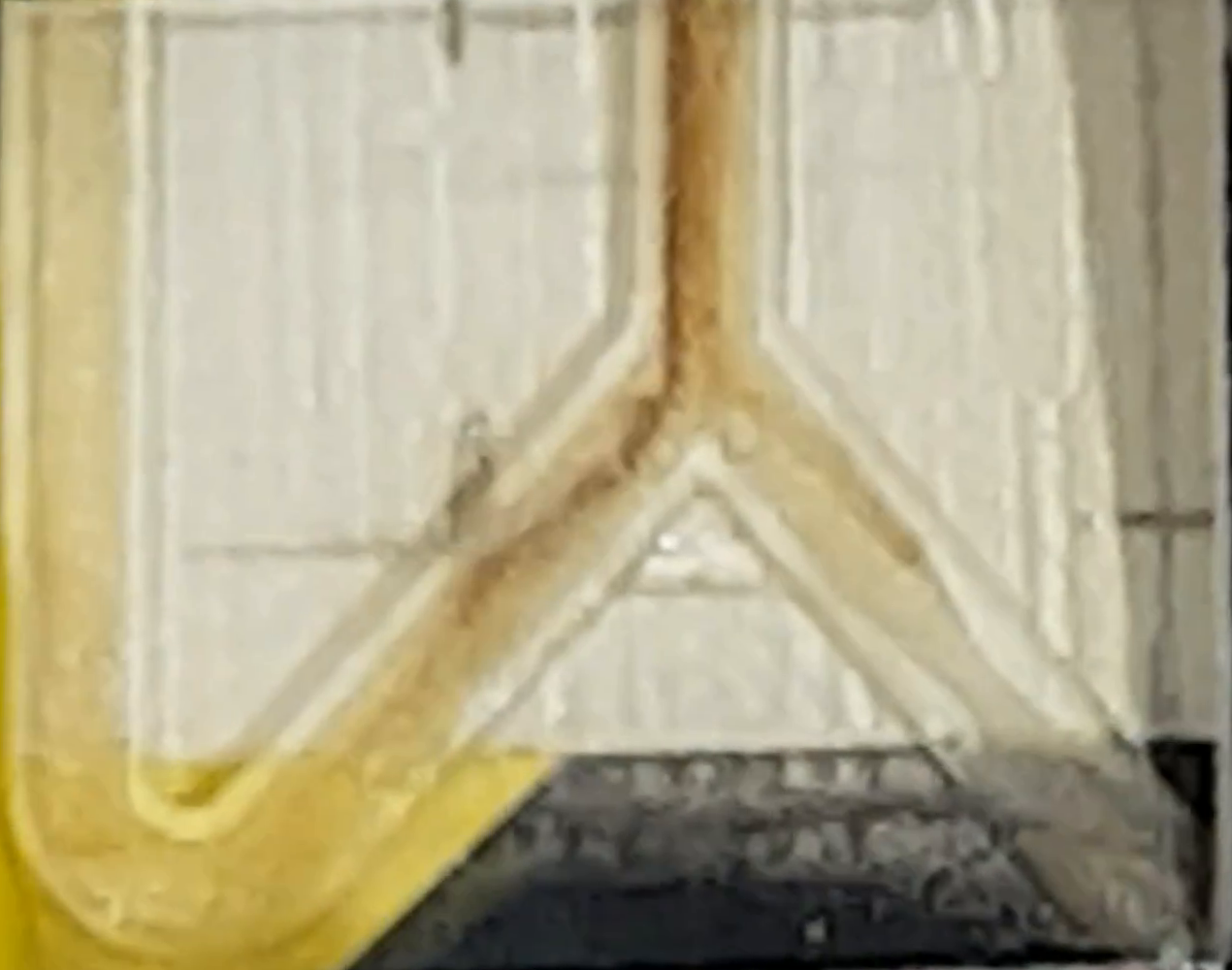}};
  \node[very thick] at (-0.75,1) {Force $F_y$=0};
  \draw[->,very thick] (1.5,0.8) node[above]{Stenosis} -- (1.4,-0.2);
   \draw[->,ultra thick,opacity=0.0] (\w3,\h3)  -- (\w3,0) node[midway,left]{x};
  \draw[->,ultra thick,opacity=0.0] (\w3,\h3)  -- (0.0,\h3) node[midway,below]{y};
  \end{tikzpicture}
  \centerline{\parbox{3.8cm}{(b) Control: $F_y=0$ 60\% stenosis, velo \SI{6.87}{\milli\liter\per\second}}}\medskip
\end{minipage}
\caption{(a) The particles are moved into the right branch containing the 60\% stenosis at flow velocity \SI{6.87}{\milli\liter\per\second}. In the control measurement at velocity \SI{6.87}{\milli\liter\per\second} most of the particles flow through the left side and only a small amount enter the stenosis.}
\label{fig:ResultExp3Stenosis60}%
\end{figure}
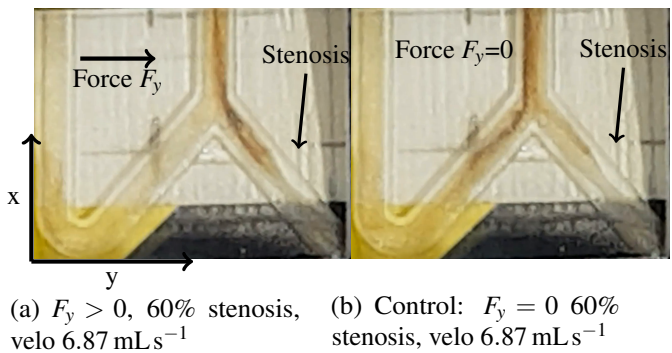

In Fig.~\ref{fig:ResultExp3Stenosis60}(a) a snapshot of the 60\% stenosis measurement shows for velocity \SI{6.87}{\milli\liter\per\second} that the force of the magnetic field pushes the particles through the bifurcation branch of the stenosis, although the flow velocity is reduced in that branch. The control measurement with no magnetic force active underlines that more particles flow through the left side with no stenosis, as seen in Fig.~\ref{fig:ResultExp3Stenosis60}(b), due to a higher flow velocity in that bifurcation branch. 

\def\w4{-2.15}
\def\h4{-1.80}
\def\lwp4{0.24}
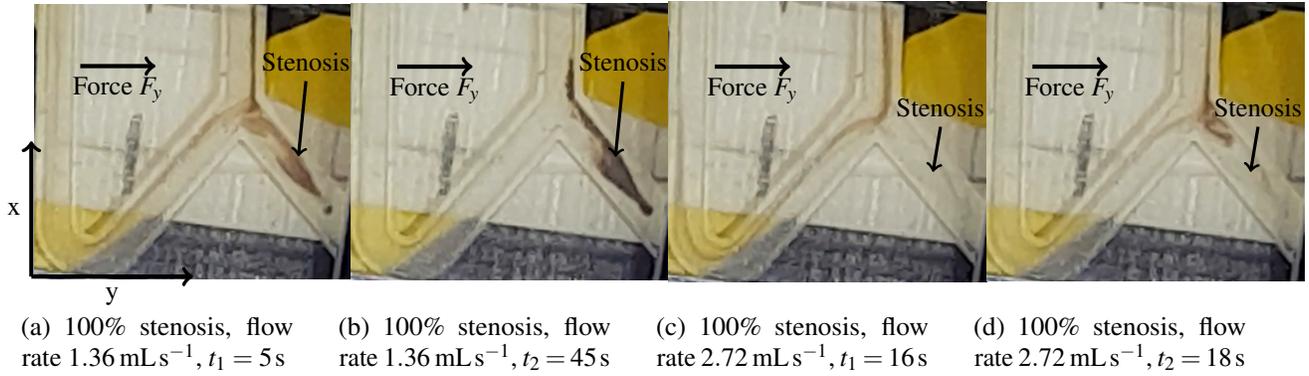
\begin{figure}[htb!]%
\begin{minipage}[b]{\lwp4\linewidth}
  \centering
  \begin{tikzpicture}
  \node at (0,0) {\includegraphics[width=1.0\linewidth]{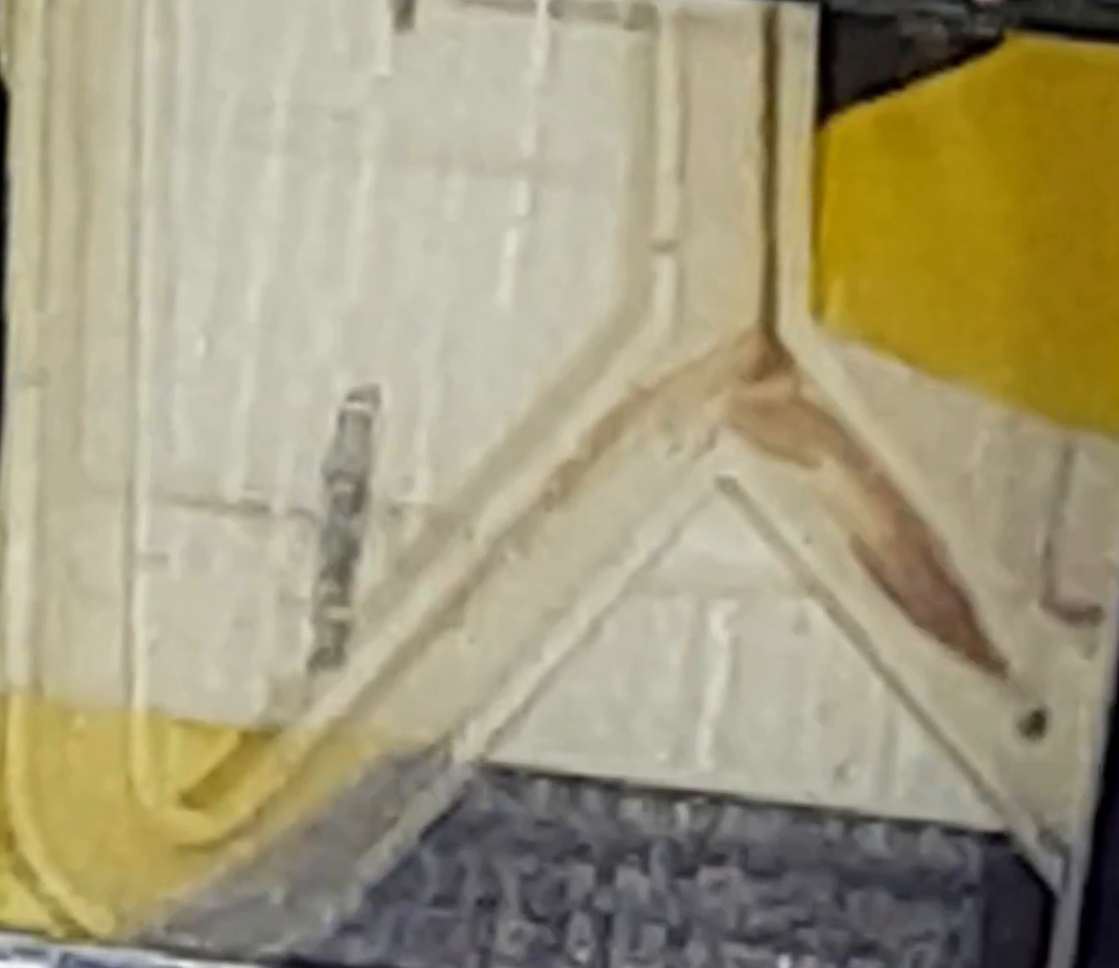}};
  \draw[->,ultra thick] (-1.5,1) -- (-0.5,1) node[midway,below]{Force $F_y$};
  \draw[->,very thick] (1.5,0.8) node[above]{Stenosis} -- (1.4,-0.2);
  \draw[->,ultra thick] (\w4,\h4)  -- (\w4,0) node[midway,left]{x};
  \draw[->,ultra thick] (\w4,\h4)  -- (0.0,\h4) node[midway,below]{y};
  \end{tikzpicture}
  \centerline{\parbox{3.6cm}{(a) 100\% stenosis, flow rate \SI{1.36}{\milli\liter\per\second}, $t_1=\SI{5}{\second}$}}\medskip
\end{minipage}%
\begin{minipage}[b]{\lwp4\linewidth}
  \begin{tikzpicture}
  \node at (0,0) {\includegraphics[width=1.0\linewidth]{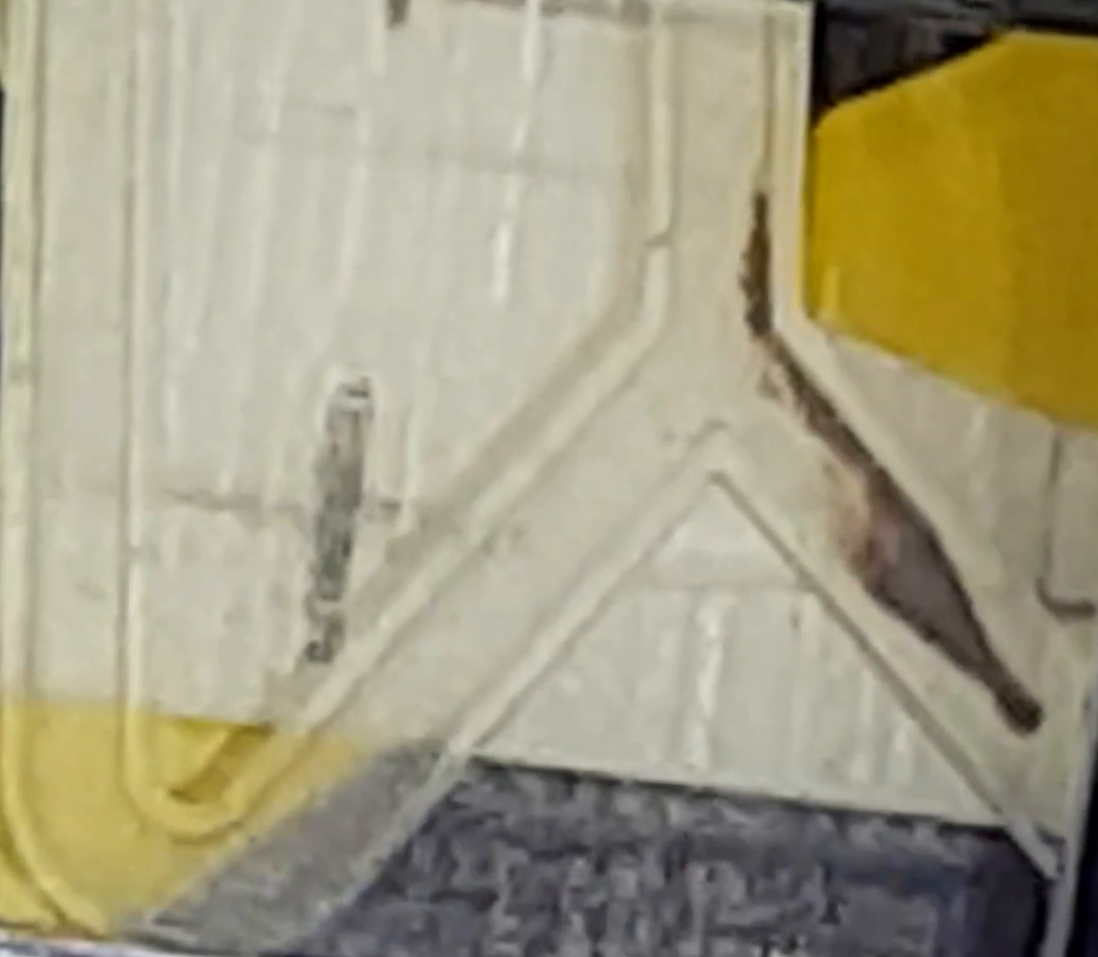}};
  \draw[->,ultra thick] (-1.5,1) -- (-0.5,1) node[midway,below]{Force $F_y$};
  \draw[->,very thick] (1.5,0.8) node[above]{Stenosis} -- (1.4,-0.2);
  \draw[->,very thick,opacity=0.0] (\w4,\h4)  -- (\w4,0) node[midway,left]{x};
  \draw[->,very thick,opacity=0.0] (\w4,\h4)  -- (0.0,\h4) node[midway,below]{y};
  \end{tikzpicture}
  \centerline{\parbox{3.6cm}{(b) 100\% stenosis, flow rate \SI{1.36}{\milli\liter\per\second}, $t_2=\SI{45}{\second}$}}\medskip
\end{minipage}%
\begin{minipage}[b]{\lwp4\linewidth}
  \centering
  \begin{tikzpicture}
  \node at (0,0) {\includegraphics[width=1.0\linewidth]{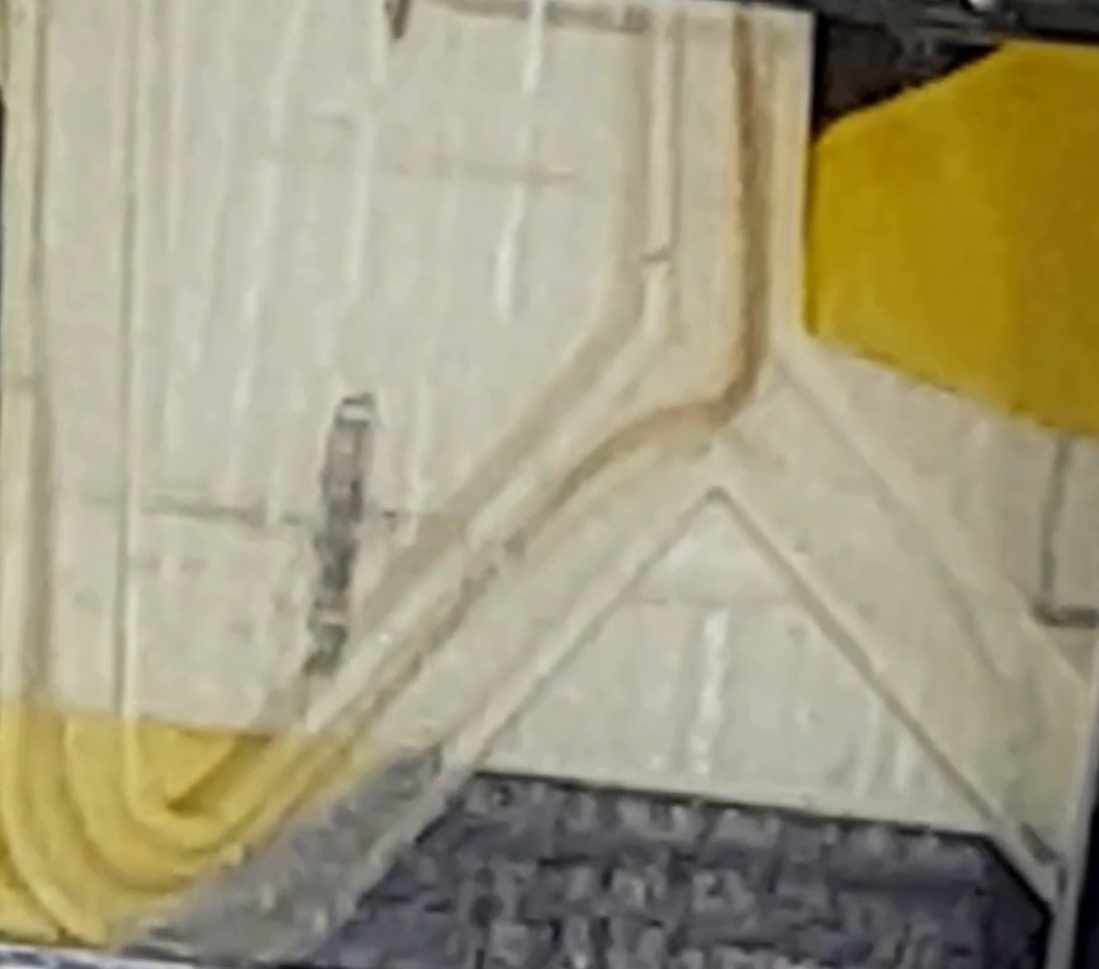}};
  \draw[->,ultra thick] (-1.5,1) -- (-0.5,1) node[midway,below]{Force $F_y$};
  \draw[->,very thick] (1.5,0.2) node[above]{Stenosis} -- (1.4,-0.4);
  \draw[->,very thick,opacity=0.0] (-2.15,-1.8)  -- (-2.15,0) node[midway,left]{x};
  \draw[->,very thick,opacity=0.0] (-2.15,-1.8)  -- (0.0,-1.8) node[midway,below]{y};
  \end{tikzpicture}
  \centerline{\parbox{3.6cm}{(c) 100\% stenosis, flow rate \SI{2.72}{\milli\liter\per\second}, $t_1=\SI{16}{\second}$}}\medskip
\end{minipage}%
\begin{minipage}[b]{\lwp4\linewidth}
  \centering
  \begin{tikzpicture}
  \node at (0,0) {\includegraphics[width=1.0\linewidth]{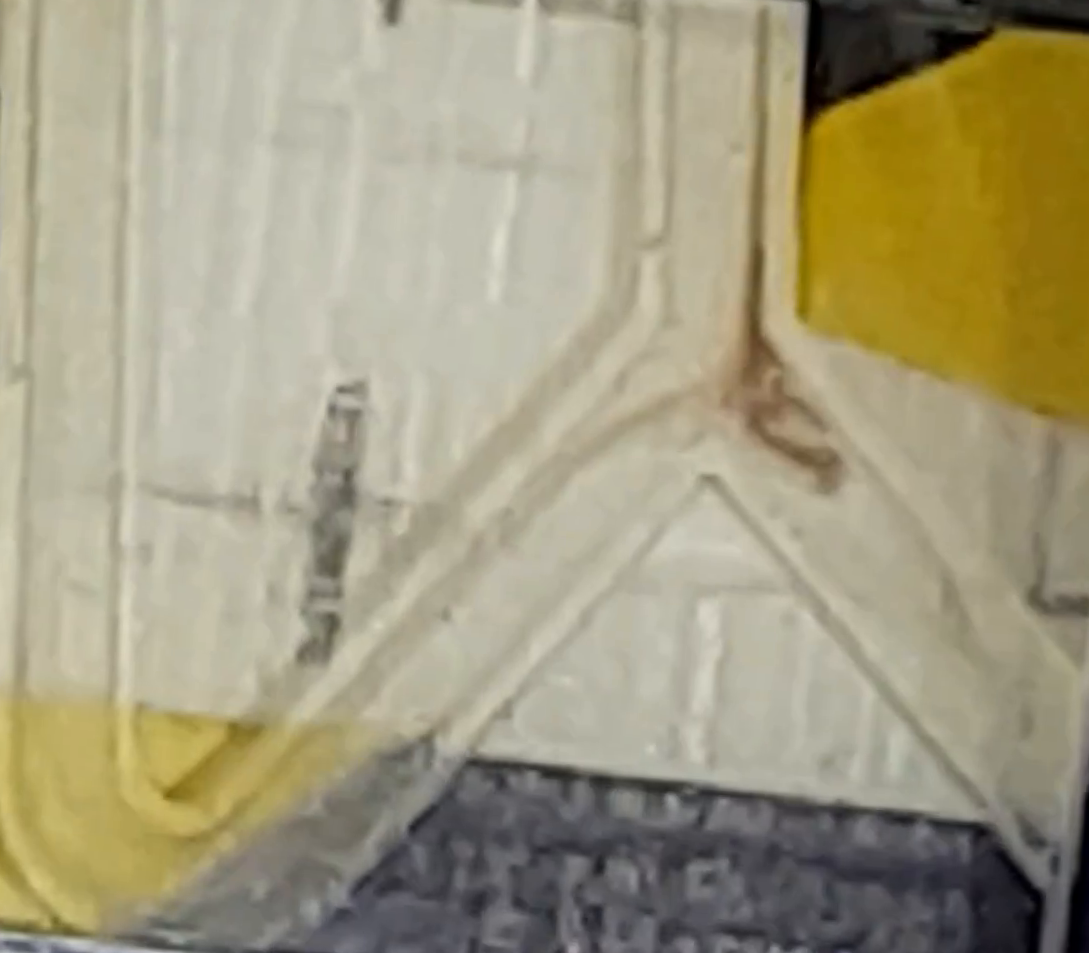}};
  \draw[->,ultra thick] (-1.5,1) -- (-0.5,1) node[midway,below]{Force $F_y$};
  \draw[->,very thick] (1.5,0.2) node[above]{Stenosis} -- (1.4,-0.4);
  \draw[->,very thick,opacity=0.0] (-2.15,-1.8)  -- (-2.15,0) node[midway,left]{x};
  \draw[->,very thick,opacity=0.0] (-2.15,-1.8)  -- (0.0,-1.8) node[midway,below]{y};
  \end{tikzpicture}
  \centerline{\parbox{3.6cm}{(d) 100\% stenosis, flow rate \SI{2.72}{\milli\liter\per\second}, $t_2=\SI{18}{\second}$}}\medskip
\end{minipage}
\caption{(a) and b)  Most of the particles accumulate in the 100\% stenosis during the flow measurements with velocity \SI{1.36}{\milli\liter\per\second}. c) and d) At a higher velocity of \SI{2.72}{\milli\liter\per\second} only a small amount of particles end up in the 100\% stenosis while most of the particles are bent towards the left side.}
\label{fig:ResultExp3Stenosis100}%
\end{figure}

Furthermore, the results of the experiments performed using the bifurcation phantom with a 100\% stenosis in the right branch are shown in Fig.~\ref{fig:ResultExp3Stenosis100}. The flow velocity is \SI{1.36}{\milli\liter\per\second} for the images in (a and b) and \SI{2.72}{\milli\liter\per\second} for the images in (c and d). In the case of the flow velocity \SI{1.36}{\milli\liter\per\second} the magnetic force of the focus field is strong enough to move the particles inside the right branch into the stenosis, although the flow velocity inside the right branch is zero due to full blockage. In the inbound tube of the phantom the particles are pushed to the far-right side but as soon as they reach the entrance of the right branch they slightly bend towards the left branch. Most of the particles still end up in the stenosis as seen in Fig.~\ref{fig:ResultExp3Stenosis100}(a,b). 
The slight bend towards the left is clearly visible in Fig.~\ref{fig:ResultExp3Stenosis100}(c,d) at flow velocity \SI{2.72}{\milli\liter\per\second}. Here, the flow dynamics bend most of the particles away from the entrance of the right branch to the left side, even though the particles are pushed to the far right in the inbound tube. Only a small amount of particles end up at the stenosis due to turbulent whirls. The video references of the 100\% stenosis measurements for the two different velocities can be found in Table \ref{tab:Exp3Stenosis100}.

\begin{table}[hbt!]
\caption{Videos results for 100\% stenosis measurements for flow velocities \SI{1.36}{\milli\liter\per\second} and \SI{2.72}{\milli\liter\per\second}.}
\label{tab:Exp3Stenosis100}
\centering
\begin{tabular}{ c c c p{9cm}  }	
\toprule
   Video ref & Flow rate inbound/ &Bifurcation & Description of results \\
   &outbound [ml/s]&&\\ \midrule
  v3.0.0R & 1.36  / 1.36  & right & Large amount of particles flow to 100\% stenosis in right branch   \\
  v3.1.0R & 2.72  / 2.72  & right & Particles flow through left branch. Only a small amount end up in the 100\% stenosis  \\
  \bottomrule
\end{tabular}
\end{table}

\subsection{Magnetic Particle Imaging}
\begin{figure}[htb!]%
\begin{minipage}[b]{0.5\linewidth}
  \centering
\begin{tikzpicture}
  \node[rotate=0](image2) at (0,0) {\includegraphics[width=0.5\linewidth]{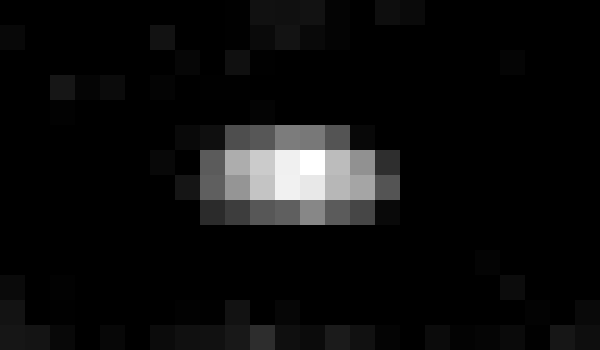}};
  \node[rotate=90,left=0pt of image2,yshift=+0.7cm,xshift=1.65cm] (profz) {\includegraphics[width=0.35\linewidth]{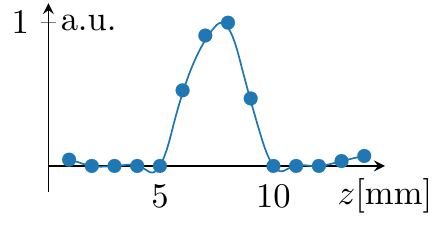}};
  \node[rotate=0,above =0pt of image2,yshift=-0.3cm] {\includegraphics[width=0.62\linewidth]{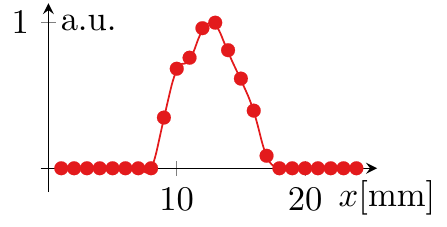}};
  \draw[c6,very thick] (-2.2,0.1)--(2.2,0.1);
  \draw[c2,very thick] (0.1,-1.25)--(0.1,1.25);
  \end{tikzpicture}
   \centerline{a)}\medskip
\end{minipage}%
\begin{minipage}[b]{0.5\linewidth}
  \centering
  \begin{tikzpicture} 
  \node[rotate=0](image1) at (0,0) {\includegraphics[width=0.5\linewidth]{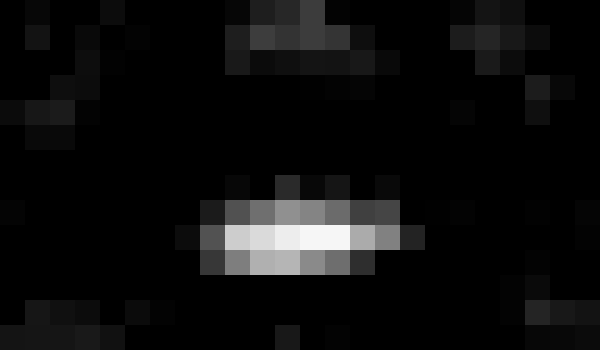}};
  \node[rotate=90,left=0pt of image1,yshift=+0.7cm,xshift=1.65cm] (profz) {\includegraphics[width=0.35\linewidth]{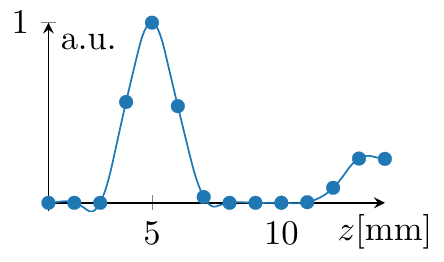}};
  \node[rotate=0,above =0pt of image1,yshift=-0.3cm] {\includegraphics[width=0.62\linewidth]{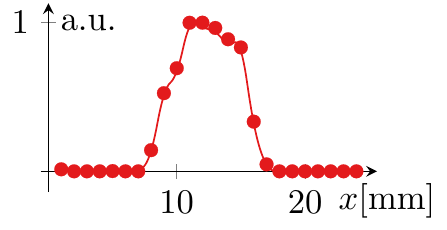}};
  \draw[c6,very thick] (-2.2,-0.4)--(2.2,-0.4);
  \draw[c2,very thick] (0.1,-1.25)--(0.1,1.25);
  \end{tikzpicture}
  \centerline{b)}\medskip
\end{minipage}%
\caption{(a) Image of a delta sample filled with beads in the $xz$-plane with profile in the $x$-direction (above) and profile in the $z$-direction (left). b) Image of delta sample containing nanomag/synomag-D 333 in the $xz$-plane with profile in the $x$-direction (above) and profile in the $z$-direction (left).}
\label{fig:DeltaSample}%
\end{figure}
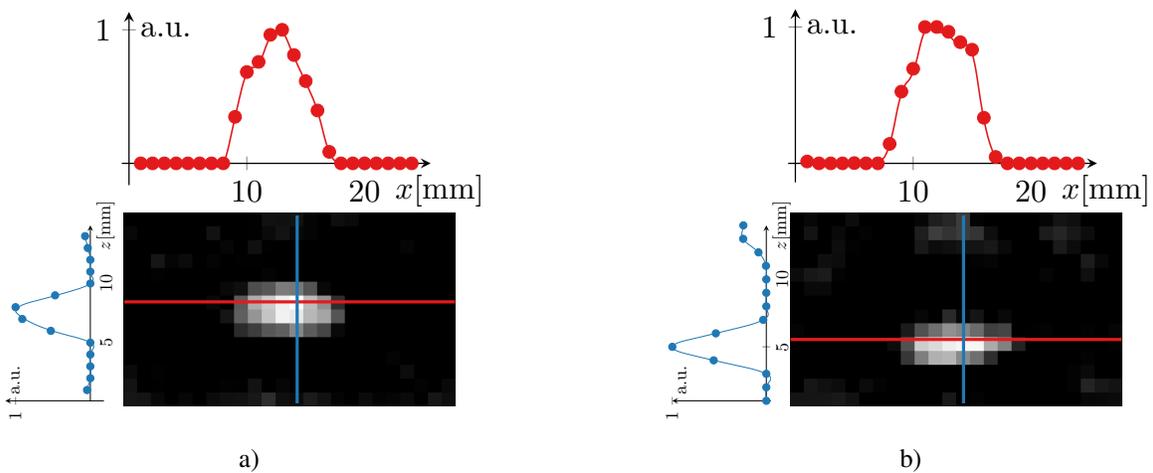

Imaging experiments without force application are performed using the most promising particles to evaluate their imaging capabilities. For these experiments, we show the reconstructed images and calculate the spatial resolution in terms of the full width at half maximum (FWHM) of the point-shaped samples in the reconstructed image.
A reconstructed image in the $xz$-plane of a static delta sample filled with nanomag/synomag-D 333 is shown in Fig.~\ref{fig:DeltaSample}(a). In order to determine the FWHM of the sample, profiles in the $x$- and $z$-directions are generated, as shown in Fig.~\ref{fig:DeltaSample}(a) (above and left of the image). The FWHM in the $x$-direction is \SI{6.7}{\milli\meter} and in the $z$-direction is \SI{2.6}{\milli\meter} indicating a better spatial resolution in the $z$-direction compared to the $x$-direction.
For the beads, the reconstructed image in the $xz$-plane is given in Fig.~\ref{fig:DeltaSample}(b) and their profiles in the $x$-direction and $z$-direction (above and left of the image) result in a FWHM of \SI{6.3}{\milli\meter} and \SI{3.0}{\milli\meter}.
Because the gradient strength in the $x$-direction and $y$-direction and DF-amplitudes are the same, the spatial resolution in the $y$-direction is assumed to be the same as in the $x$-direction.

\subsection{Magnetic Particle Imaging and Navigation in Flow}
In this bifurcation experiment with 100\% stenosis at velocity \SI{1.36}{\milli\liter\per\second} the imaging and navigation mode of MPI is successfully used to navigate the particles towards the stenosis while the distribution of the particles is captured with the imaging mode. 
With a ratio between $\zeta=20$ and $\varphi=1$ for force and imaging mode, the induced magnetic force acts sufficiently long enough to maneuver the particles towards the stenosis, although no force is acting on the particles during the short time of the imaging mode. 
\begin{figure}[htb!]%
\begin{minipage}[b]{0.6\linewidth}
  \centering
  \begin{tikzpicture}
  \node at (0,0) {\includegraphics[width=1.0\linewidth]{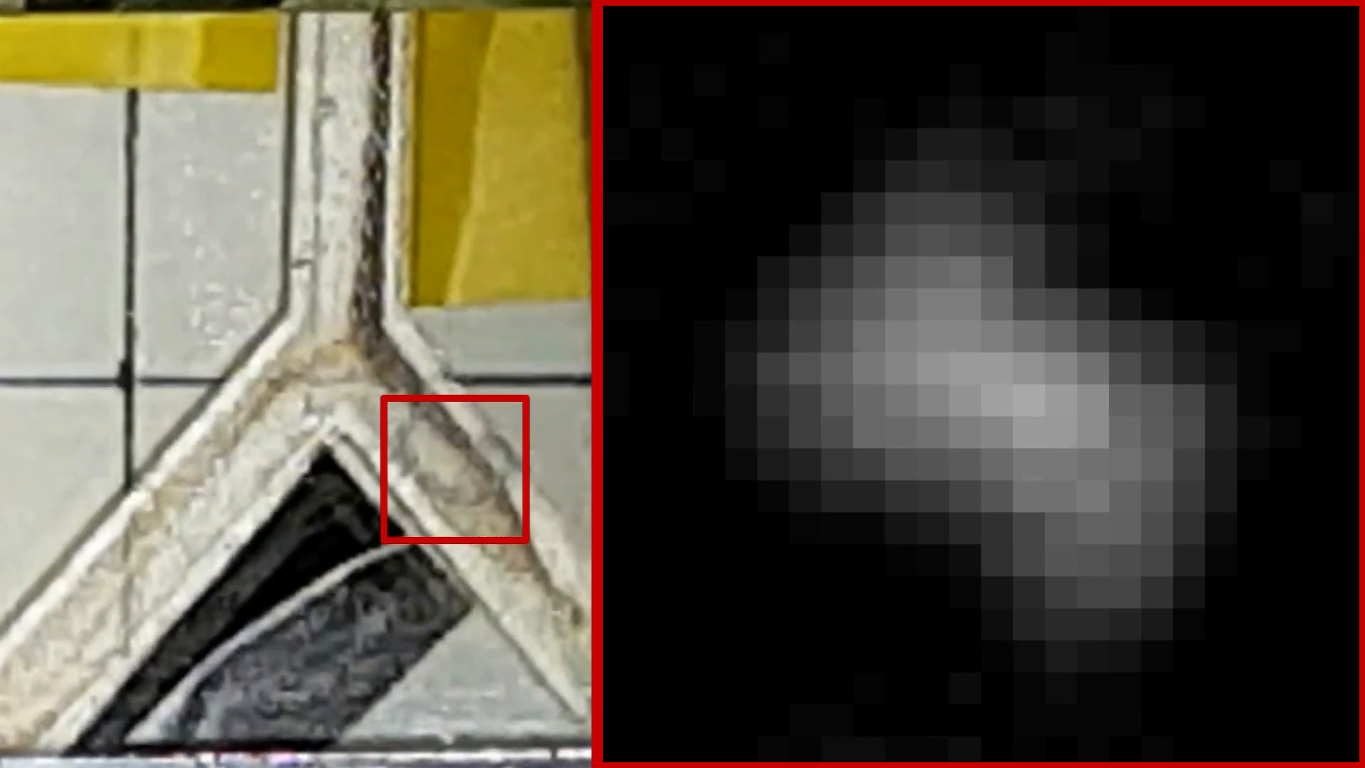}};
  \draw[->,ultra thick] (-4.5,1) -- (-3.5,1) node[midway,below]{Force $F_y$};
  \draw[->,very thick] (-1.5,0.8) node[above]{Stenosis} -- (-1.4,-0.2);
  \draw[->,ultra thick] (-1.0\linewidth/2,-3)  -- (-1.0\linewidth/2,0) node[midway,left]{x};
  \draw[->,ultra thick] (-1.0\linewidth/2,-3)  -- (-1.0\linewidth/4,-3) node[midway,below]{y};
  \end{tikzpicture}
  \centerline{(a) 100\% stenosis, flow rate \SI{1.36}{\milli\liter\per\second}, MPI image }\medskip
\end{minipage}\hspace{1.0mm}%
\begin{minipage}[b]{0.37\linewidth}
  \begin{tikzpicture}
  \node at (0,-0.05) {\includegraphics[width=1.0\linewidth]{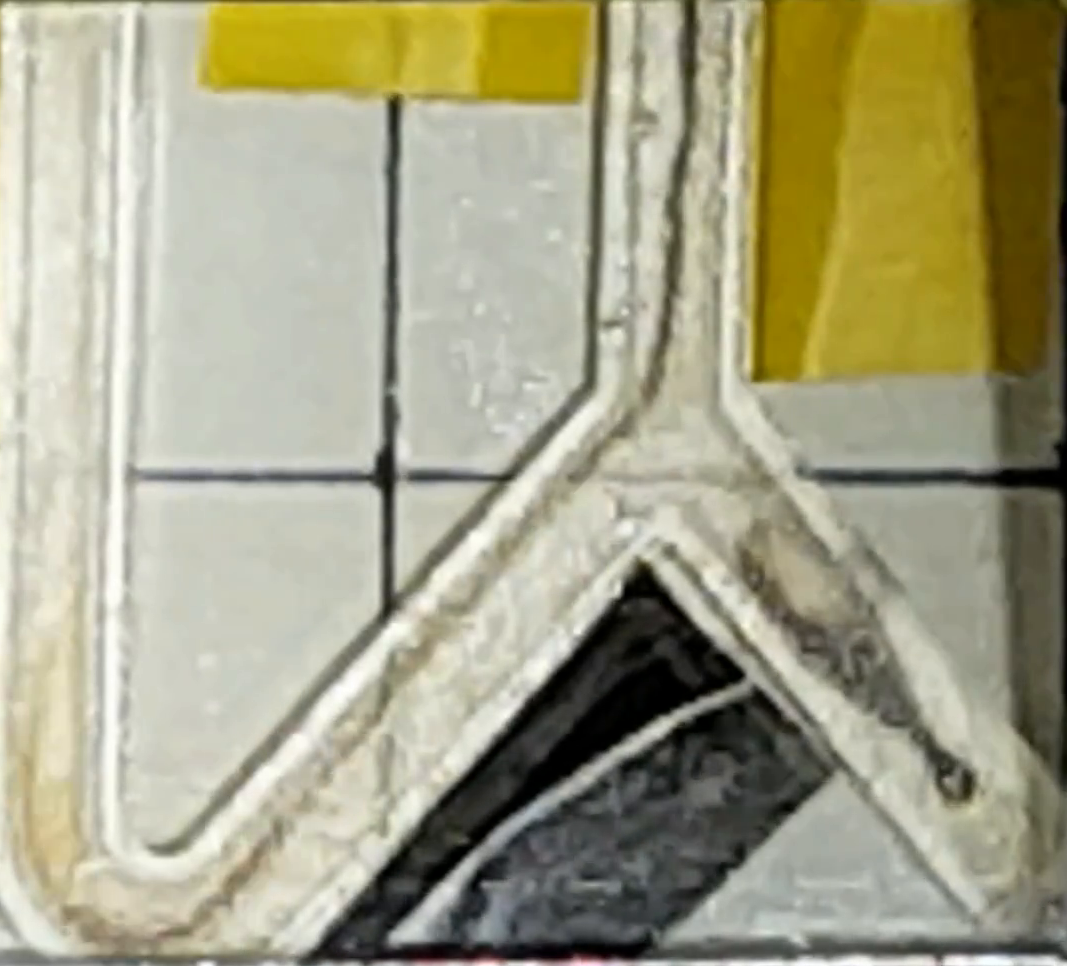}};
  \node[very thick] at (-1.25,1) {$F_y$=0};
  \draw[->,very thick] (2.5,0.8) node[above]{Stenosis} -- (2.4,-0.2);
  \draw[->,ultra thick,opacity=0.0] (-1.0\linewidth/2,-3)  -- (-1.0\linewidth/2,0) node[midway,left]{x};
  \draw[->,ultra thick,opacity=0.0] (-1.0\linewidth/2,-3)  -- (0.0,-3) node[midway,below]{y};
  \end{tikzpicture}
  \centerline{(b) $F_y$=0, 100\% stenosis, flow rate \SI{1.36}{\milli\liter\per\second} }\medskip
\end{minipage}
\caption{(a) Most of the particles are moved into the right side towards the 100\% stenosis at flow velocity \SI{1.36}{\milli\liter\per\second} while the MPI imaging mode shows the distribution of the particles. (b) In the control measurement afterwards where no forces, where applied all particles flow through the left side. }
\label{fig:ResultExp3MPIFStenosis100}%
\end{figure}
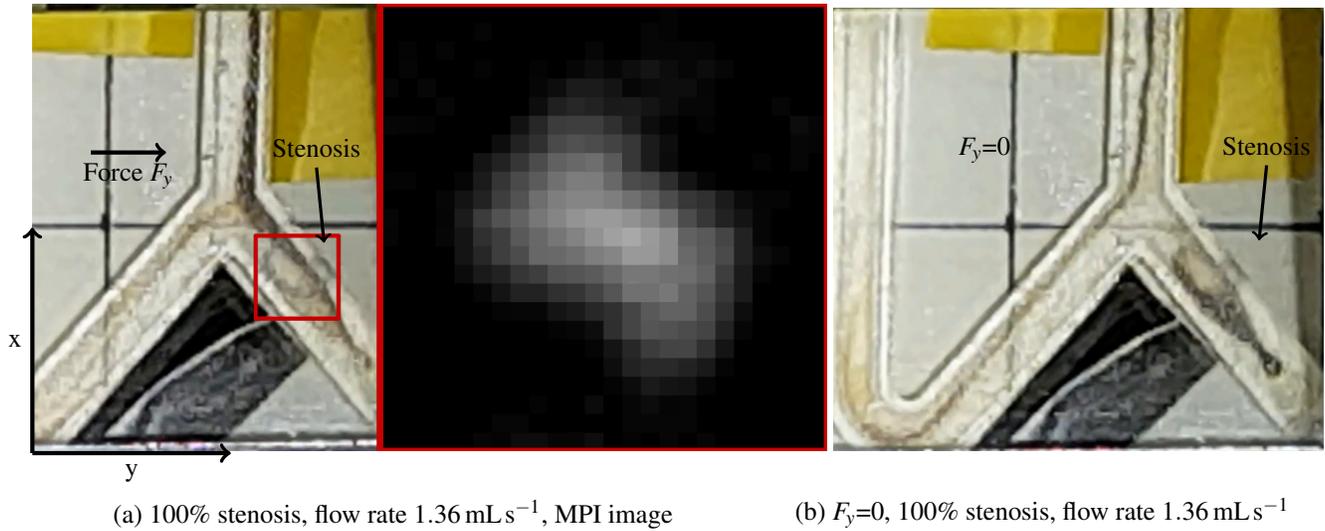
The particles are driven towards the right branch of the bifurcation, while the intensities within the MPI image increase as the particles enter the right branch and are pushed to the stenosis, as seen in Fig.~\ref{fig:ResultExp3MPIFStenosis100}(a). In the control measurement afterwards, shown in Fig.~\ref{fig:ResultExp3MPIFStenosis100}(b), the particles are flowing completely through the left branch of the bifurcation. The references of the videos can be found in Table~\ref{tab:Exp3Stenosis100MPI}.
\begin{table}[hbt!]
\caption{Video results for 100\% stenosis measurements for flow velocities \SI{1.36}{\milli\liter\per\second} while taking snapshots within imaging mode.}
\label{tab:Exp3Stenosis100MPI}
\centering
\begin{tabular}{ c c c p{9cm}  }	
\toprule
   Video ref & Flow rate inbound/ &Bifurcation & Description of results \\
   &outbound [ml/s]& & \\ \midrule
  v4.0.0R & 1.36  / 1.36  & right & Large amount of particles flow to 100\% stenosis in right branch   \\
  v4.0.1R & 1.36  / 1.36  & right & Control: With no force, all particles flow through left branch  \\
  \bottomrule
\end{tabular}
\end{table}

\section{Discussion}
The results from the MPS measurements show that nanomag/synomag-D 333 particles provide the best compromise between imaging capabilities and magnetophoretic mobility of all investigated nanomag/synomag-D particles. The spectrum of nanomag/synomag-D 333 particles indicates a sufficient MPI imaging performance with about 35 harmonics above the noise level. The performance of the Dynabeads MyOne particles is slightly inferior but they generate sufficient harmonics for MPI imaging. The imaging characteristics, in terms of the spectra of both particle types, are not influenced by magnetizing or vertexing them. Thus, both particle types do not show a remaining remanence. For the nanomag/synomag-D 333 particles the intensities of the first five frequencies show a strong linearity dependent on the particle concentration. The regression lines have coefficients of determination of $R=0.996$ to $R=0.997$. This strong linearity is inevitable for MPI applications since the reconstruction principle requires linearity. This criterion is also met by the DynaBeads MyOne particles where intensities of the frequency signals indicate a linear tendency to the iron concentration, but their coefficients of determination for the regression lines are smaller at $R=0.971$ to $R=0.986$. 
 The similar half separation times of the DynaBeads and nanomag/synomag-D batches (333,334) particles indicate an equal magnetophoretic mobility for both particle types. The nanomag/synomag-D 333 batch is therefore suitable for magnetic navigation and has shown the most promising imaging characteristics for MPI by generating the largest number of harmonics above the noise level.
 Additionally, the investigated polystyrene Dynabeads MyOne particles are not biodegradable, thus are not suitable for human use. The nanomag/synomag-D particles do have a biocompatible coating layer and are therefore much more suited to being resolved by the liver. Due to the better performance in imaging characteristics, similar half time separation and biocompatibility, the nanomag/synomag-D 333 particles are used for further MPN bifurcation flow and MPIN flow bifurcation experiments.

The results of the Magnetic Particle Navigation experiments demonstrate that nanomag/synomag-D 333 particles can be actuated to one side of the bifurcation junction within a liquid medium flowing with a velocity of up to \SI{8.18}{\milli\liter\per\second} (\SI{652.1}{\milli\meter\per\second}). The rectangular cross-section $A_{\text{exp}}=(\SI{3.544}{\milli\meter})^2=\SI{12.5}{\square\milli\meter}$ of the tube corresponds to a cross-section of circular vessel tube with diameter of \SI{4}{\milli\meter}. In the literature \cite{ford_characterization_2005} the flow rate \SI{4.58}{\milli\liter\per\second} (\SI{233.2}{\milli\meter\per\second}) within the internal carotid artery with circular diameter of about \SI{5}{\milli\meter}\cite{krejza_carotid_2006} is stated. Thus, it is promising to investigate Magnetic Particle Navigation with MPI under realistic conditions within blood circulation.

The effect of the magnetic navigation could be slightly stronger in a real heart since the peristaltic pump generates a constant volume flow whereas the heart pumps with a frequency of about \SI{60}{\hertz}. During diastole the flow volume is noticeably lower, which would give the magnetic force more time to move the particles in the desired direction. With this in mind, it might be possible to move particles from the aorta into the common carotid artery.

For flow velocity \SI{2.72}{\milli\liter\per\second} (\SI{217}{\milli\meter\per\second}) the results show that it is even possible to stop particles by pressing them against the side wall of the phantom. The friction at the side wall and the reduced flow velocity profile close to the side wall are probably responsible for this effect. These findings indicate a promising prospect for targeted drug delivery applications in small arteries where the effect of the drug could be enhanced. The effect could be further investigated by trying to stop particles explicitly with the magnetic force against the flow direction.

Furthermore, the flow bifurcation experiment with a stenosis of 60\% show that the nanomag/synomag-D 333 particles can be pushed to one side of a bifurcation junction up to a flow velocity of \SI{6.87}{\milli\liter\per\second} (\SI{546.6}{\milli\meter\per\second}). The 60\% stenosis increases the flow resistance in its bifurcation branch and a larger amount of the flow volume prefers to flow through the clear side. These circumstances make it more challenging to move the particles through the stenosis. 
If the stenosis in one branch is a full blockage of 100\%, it is still possible to magnetically force the particles to the stenosis at a maximal flow velocity of \SI{1.36}{\milli\liter\per\second} (\SI{108.2}{\milli\meter\per\second}). With these findings it might be possible to resolve a stenosis caused by blood clots in a medium-sized artery by using drug-loaded particles to clear blockages. Under these conditions a drug normally injected intravenously would never reach the stenosis. Thus, even a small amount of particles loaded with tissue plasminogen activator (tPA) reaching the blood clot at the stenosis would be highly benefical.

Both particle types show that their MPI signal is sufficient enough to reconstruct an image. But the spatial resolutions, in terms of FWHM in the $x$-direction at \SI{6.7}{\milli\meter} and \SI{2.6}{\milli\meter} in the  $z$-direction for Dynabeads MyOne particles, are not as good as dedicated imaging particles. Rahmer et al.\cite{rahmer_analysis_2012} achieved spatial resolutions in the sub-millimeter range. For nanomag/synomag-D 333 particles the spatial resolutions in terms of FWHM in the $x$-direction at \SI{6.3}{\milli\meter} and \SI{3.0}{\milli\meter} in the $z$-direction, are similar to the Dynabeads MyOne particles.

Finally, the results of the Magnetic Particle Imaging and Navigation experiments demonstrate that it is possible to control and image particles through a bifurcation at a flow velocity of \SI{1.36}{\milli\liter\per\second} (\SI{108.2}{\milli\meter\per\second}) quasi-simultaneously. The particles are navigated through a bifurcation towards the stenosis branch even when 100\% blocked.

As seen in video v4.0.0R, it is challenging to maneuver the particles to towards the 100\% stenosis since the total blockage creates a high resistance. At the boundary surface of the bifurcation turbulances occur and the negative pressure caused by the flowing liquid pulls the particles out of the stenosis back into the flowing current. With a ratio of 20 to 1 between navigation and imaging, an imaging rate of \SI{2.15}{\second} per image can be achieved to identify the position of the particles.

In terms of improvement, the limiting flow velocity of \SI{1.36}{\milli\liter\per\second} (\SI{108.2}{\milli\meter\per\second}) could be increased by stronger focus fields to induce a greater force on the particles. Such an adaption is not easily possible with the commerical system used in this work. But with a custom built human scanner\cite{Rahmer2017SciRob,graeser_human-sized_2019} it would be feasible to introduce stronger additional focus fields for human brain applications with MPIN.

\section{Conclusion}
In this work, we have determined that the nanomag/synomag-D 333 particles provide the best compromise between magnetic manipulability and imaging performance for MPI. With these particles we have further demonstrated the feasibility to maneuver particles within a volume flow of \SI{8.18}{\milli\liter\per\second} (\SI{652.1}{\milli\meter\per\second}) towards one side of a bifurcation by using the MPN method of an MPI scanner. Additionally, the MPIN method has been successfully used to navigate particles towards a 100\% stenosis within a bifurcation, while imaging the particles' distribution in the stenosis every \SI{2.15}{\second} at a flow velocity of \SI{1.36}{\milli\liter\per\second} (\SI{108.2}{\milli\meter\per\second}). In the future, magnetic particles combined with a tissue plasminogen activator (tPA) might be used to resolve blood clots in hard-to-reach positions while using MPIN to monitor the liquidation of the stenosis. 

\bibliography{sample}

\section{Acknowledgements}
F.G., N.G., F.T. and T.K. thankfully acknowledge the financial support of the German Research Foundation (DFG, grant number KN 1108/2-1) and the Federal Ministry of Education and Research (BMBF, grant number 05M16GKA). This work was also supported by the BMBF under the frame of EuroNanoMed III (grant number: 13XP5060B, T.K, P.L.).

\end{document}